\documentclass[11pt,a4paper]{article}
\pdfoutput=1
\usepackage{jheppub}
\usepackage[utf8]{inputenc}

\title{Effective-field theory analysis of the \boldmath $\tau^-\to \pi^-\pi^0\nu_\tau$ decays}

\author[1]{J. A. Miranda}
\author[1]{and P. Roig}

\affiliation[1]{Departamento de F\'isica, Centro de Investigaci\'on y de Estudios Avanzados del IPN,\\
Apdo. Postal 14-740,07000 Ciudad de M\'exico, M\'exico.}

\emailAdd{jmiranda@fis.cinvestav.mx}
\emailAdd{proig@fis.cinvestav.mx}

\abstract{We perform an effective field theory analysis of the $\tau^- \to \pi^- \pi^0 \nu_\tau$ decays, that includes the most general interactions between Standard Model fields up to dimension six, assuming left-handed neutrinos. We constrain as much as possible the necessary Standard Model hadronic input using chiral symmetry, dispersion relations, data and asymptotic QCD properties. As a result, we set precise (competitive with low-energy and LHC measurements) bounds on (non-standard) charged current tensor interactions, finding a very small preference  for their presence, according to Belle data. Belle-II near future measurements can thus be very useful in either confirming or further restricting new physics tensor current contributions to these decays. For this, the spectrum in the di-pion invariant mass turns out to be particularly promising. Distributions in the angle defined by the $\tau^-$ and $\pi^-$ momenta can also be helpful if measured with less than $10\%$ accuracy, both for non-standard scalar and tensor interactions.}

\notoc
\begin{document}
\maketitle
\flushbottom
\section{INTRODUCTION}
Early studies of nuclear beta decays and, particularly, the problem of apparent non-conservation of energy and violation of the spin-statistics theorem lead to Pauli's postulation of the neutrino. Soon after, Fermi proposed a theory \cite{Fermi:1934hr} describing these decays which was inspired by QED's vector current interaction which, however, was of a local current-current type. This was the first step towards establishing the V-A nature of the weak force and understanding its maximal parity violation. Now the original Fermi theory is regarded as one of the possible contributions of dimension six effective operators to these decays and it constitutes the basis for effective field theories. In this spirit, not only nuclear beta decays, but also purely leptonic lepton decays, pion decays into a lepton and its corresponding neutrino and also strangeness-changing meson and baryon decays involving a lepton charged current can be studied in a coherent and comprehensive way with direct connection to the underlying theory at some TeVs \cite{a25, a24, Cirigliano:2013xha, Ciri3, a26, Courtoy:2015haa, a27, a28, Alioli:2017ces, Gonzalez-Alonso:2017iyc, Cirigliano:2018dyk}. Thus, it is possible to obtain bounds on non-standard charged current interactions from either of these processes that can be compared among them (assuming lepton universality if necessary). As a result, quite generic New Physics (NP) is restricted in absence of deviations from the Standard Model (SM) predictions. In the event of any such departures appearing, one would expect them to point to the underlying new dynamics, as (nuclear) beta and muon decays did with the $W$ mass value (provided the coupling intensity can be estimated from some symmetry argument) and its left-handed couplings.

In ref. \cite{E} we put forward that semileptonic tau decays are also an interesting scenario in this respect. Particularly, our study of the $\tau^- \to \pi^- (\eta/\eta') \nu_\tau$ decays \cite{E} showed that they could be competitive with superallowed nuclear beta decays in restricting scalar non-standard interactions. Our aim in this paper is to extend our previous analysis to the $\tau^- \to \pi^- \pi^0 \nu_\tau$ decays, which should not be sensitive to NP charged current scalar interactions (as generally, they are very suppressed by the small isospin breaking effects giving rise to them in this decay channel \cite{Cirigliano:2001er}) but could instead be very competitive restricting charged-current tensor interactions. The recent letter \cite{Cirigliano:2018dyk} also addresses this question.

Only if the SM input (and particularly the hadronization) to the considered decays is well under control one can actually set bounds on NP effective couplings. This is the case for the vector and -to a lesser extent- the scalar interactions (where we will follow the treatment in refs. \cite{Descotes} and \cite{Dumm}, respectively) but only a theory-driven approach is possible for the tensor form factor (where we will complement our previous work \cite{E} guided by refs. \cite{Cirigliano:2017tqn} and \cite{Baum:2011rm}). In all cases it is desirable to fulfill the requirements imposed by the approximate chiral symmetry of QCD, which are automatically enforced in its low-energy effective field theory, Chiral Perturbation Theory ($\chi PT$) \cite{ChPT1, ChPT2, ChPT3}. If possible, it is also convenient to use dispersion relations to warrant analyticity and comply with unitarity, at least in the elastic region (for the $\pi\pi$ system it amounts to $\sim 1$ GeV). Within this formalism, known short-distance QCD constraints \cite{Brodsky:1973kr, Lepage:1980fj} can also be satisfied. In the absence of data (as it the case for the tensor form factor) enlarging the domain of applicability of $\chi PT$ coupled to tensor sources \cite{Mateu:2007tr, Mateu} by including resonances as explicit degrees of freedom \cite{Ecker, Ecker:1989yg} could seem useful, although we will show in the appendix of this paper that it is not the case.

This work is organized as follows: in section \ref{sec2} we present the basics for an effective field theory treatment of the considered decays. In section \ref{sec3} the different contributions to the matrix element are identified and the participant meson form factors defined. These are the subject of section \ref{sec4}, with a special focus on the tensor form factor%, which is first derived beyond leading order $\chi PT$ in this paper
. With all SM contributions fixed, we perform a phenomenological study in search for NP signatures, examining the hadron spectrum and branching ratio, the Dalitz plot distributions and the forward-backward asymmetry in section \ref{sec5}. The conclusions of this research are summarized in section \ref{sec6}.

\section{EFFECTIVE THEORY ANALYSIS OF $\tau^-\to\nu_\tau\bar{u}d$}\label{sec2}
For low-energy charged current processes, the effective Lagrangian with $SU(2)\otimes U(1)$ invariant dimension six operators~\footnote{See in refs.~\cite{a22,a23} the most 
general effective Lagrangian including SM fields.} reads \cite{a24,a25}
\begin{equation}
\mathcal{L}^{(eff)}=\mathcal{L}_{SM}+\frac{1}{\Lambda^2}\sum_i \alpha_i O_i\quad \to \quad\mathcal{L}_{SM}+\frac{1}{v^2}\sum_i \hat{\alpha}_i O_i,
\end{equation}
with $\hat{\alpha}_i=(v^2/\Lambda^2)\alpha_i$ the dimensionless multi-TeV NP couplings.\\
If we particularize it for the $\mathcal{O}$(1 GeV) semileptonic strangeness and lepton-flavor conserving \footnote{An EFT framework study of strangeness-changing processes 
is carried out in refs. \cite{a26,a27,a28}.} charged current transitions involving any lepton ($\ell=e,\,\mu,\,\tau$) and only left-handed neutrino fields, 
the following Lagrangian is obtained (where subscripts L(R) stand for left-handedness (right-handedness))
\begin{equation}\label{eqn:1}\begin{split}
\mathcal{L}_{CC}&=-\frac{4G_F}{\sqrt{2}}\bigl[(1+[v_L]_{\ell\ell})\bar{\ell}_L\gamma_\mu \nu_{\ell L}\,\bar{u}_L\gamma^\mu d_L+[v_R]_{\ell\ell}\,\bar{\ell}_L\gamma_\mu \nu_{\ell L}\,\bar{u}_R\gamma^\mu d_R\\
&\quad+[s_L]_{\ell\ell}\,\bar{\ell}_R \nu_{\ell L}\,\bar{u}_R d_L+[s_R]_{\ell\ell}\,\bar{\ell}_R \nu_{\ell L}\,\bar{u}_L d_R\\
&\quad+[t_L]_{\ell\ell}\,\bar{\ell}_R\sigma_{\mu\nu} \nu_{\ell L}\,\bar{u}_R\sigma^{\mu\nu} d_L\bigr]+h.c..
\end{split}\end{equation}

In the previous equation $G_F$ is the tree-level definition of the Fermi constant and $\sigma^{\mu\nu}\equiv i\left[\gamma^\mu,\gamma^\nu\right]/2$. The SM Lagrangian is recovered setting $v_L=v_R=s_L=s_R=t_L=0$. Heavy degrees of freedom (H, $W^{\pm}$ and $Z$ bosons plus $c$, $b$ and $t$ quarks) have been integrated out to obtain eq.~(\ref{eqn:1}). The effective couplings $v_{L,R}$, $s_{L,R}$ and $t_L$ generated by the NP can be taken real since we are only interested in $CP$ conserving quantities ~\footnote{Appendix A in ref. \cite{a24} provides with these couplings as functions of the $\hat{\alpha}_i$ couplings.}.

Although observables are renormalization scale and scheme independent, this scale independence comes after the cancellation of the scale dependence of the effective couplings ($v_{L,R}$, $s_{L,R}$ and $t_L$) by the corresponding scale dependence of the hadronic matrix elements. These encode the amplitude for the quark current to produce/annihilate the measured hadrons. As it conventional, we select $\mu=2\, \mathrm{GeV}$ as the renormalization scale.%$\overline{MS}$. 

It is advantageous to shift our basis for the spin-zero currents so that the new ones have defined parity. This is achieved by means of introducing $\epsilon_S=s_L+s_R$ and $\epsilon_P=s_L-s_R$. Although the other elements in the basis of currents remain unmodified, we also rename them to avoid any confusion between both bases: 
$\epsilon_{R,L}=v_{L,R}$ and $\epsilon_T=t_L$.

One can proceed with $\ell=e,\,\mu,\,\tau$ in full generality (which may be profitable if lepton universality is an approximate symmetry). We, however, focus now on the tau case (and omit the corresponding flavor subindex in the following), in such a way that the corresponding semileptonic effective Lagrangian is:
\begin{equation}\begin{split}
\mathcal{L}_{CC}&=-\frac{G_F}{\sqrt{2}}V_{ud}(1+\epsilon_L+\epsilon_R)\lbrace\bar{\tau}\gamma_\mu(1 -\gamma^5)\nu_{\tau}\,\bar{u} \big[\gamma^\mu-(1-2\hat{\epsilon}_R)\gamma^\mu\gamma^5\bigr]d\\
&\quad+\bar{\tau}(1-\gamma^5) \nu_{\tau}\,\bar{u}(\hat{\epsilon}_S -\hat{\epsilon}_P\gamma^5)d\\
&\quad+2\hat{\epsilon}_T\bar{\tau}\sigma_{\mu\nu}(1-\gamma^5) \nu_{\tau}\,\bar{u}\sigma^{\mu\nu}  d\rbrace+h.c.,
\end{split}\end{equation}
where $\hat{\epsilon}_i\equiv \epsilon_i/(1+\epsilon_L+\epsilon_R)$ for $i=R,S,P,T$. From this expression it is easily seen that, working at linear order in the $\hat{\epsilon}_i$, one is insensitive to non-standard spin-one charged current interactions because the overall dependence on $\epsilon_L+\epsilon_R$ cannot be isolated, as it is subsumed in the determination of $G_F$. That is, conveniently normalized rates cancel the overall factor  $(1+\epsilon_L+\epsilon_R)$ in the previous equation. We note that, at linear order in the $\hat{\epsilon}_i$'s, these agree with ref. \cite{a24}.

\section{SEMILEPTONIC $\tau$ DECAY AMPLITUDE}\label{sec3}
From now on, we will study the semileptonic $\tau^- \to \pi^-(P_{\pi^-})\, \pi^0(P_{\pi^0})\, \nu_\tau (P')$ decays, where pions parity determines that only scalar, vector and tensor currents contribute. The decay amplitude reads\footnote{As in ref.~\cite{E}, we take the short-distance electroweak radiative corrections encoded in $S_{EW}$ 
\cite{a29,a30,a31,a32,a33,a34,Braaten:1990ef,a35} as a global factor in eq. (\ref{SDA:eq1}). Although $S_{EW}$ does not affect the scalar and tensor contributions, the error of this approximation 
is negligible and renders simpler expressions than proceeding otherwise.} 
\begin{equation}\begin{split}\label{SDA:eq1}
\mathcal{M}&=\mathcal{M}_V + \mathcal{M}_S + \mathcal{M}_T\\
&=\frac{G_F V_{ud} \sqrt{S_{EW}}}{\sqrt{2}}(1+\epsilon_L+\epsilon_R)\bigl[L_\mu H^\mu+\hat{\epsilon}_S LH+2\hat{\epsilon}_T L_{\mu\nu}H^{\mu\nu}\bigr],
\end{split}\end{equation}
where the following lepton currents were introduced:
\begin{subequations}
\begin{align}
L_\mu&=\bar{u} (P')\gamma^\mu (1-\gamma^5)u(P),\\
L&=\bar{u} (P')(1+\gamma^5)u(P),\\
L_{\mu\nu}&=\bar{u} (P')\sigma_{\mu\nu} (1+\gamma^5)u(P).
\end{align}
\end{subequations}
The scalar ($H$), vector ($H^\mu$) and tensor ($H^{\mu\nu}$) hadron matrix elements entering eq. (\ref{SDA:eq1}) can be decomposed using Lorentz invariance and discrete QCD symmetries in terms of a number of allowed Lorentz structures times the corresponding form factors, which are scalar functions encoding the hadronization procedure. Specifically, these are
\begin{subequations}
\begin{align}
H&=\langle \pi^0\pi^- \vert \bar{d} u \vert 0\rangle\equiv F_S(s),\label{ad:2}\\
H^\mu&=\langle \pi^0\pi^- \vert \bar{d}\gamma^\mu u \vert 0\rangle=C_V Q^\mu F_+(s) + C_S \left(\frac{\Delta_{\pi^-\pi^0}}{s}\right)q^\mu F_0(s),\label{ad:1}\\
H^{\mu\nu}&=\langle \pi^0\pi^- \vert \bar{d}\sigma^{\mu\nu} u \vert 0\rangle=iF_T(s)(P^\mu_{\pi^0}P^\nu_{\pi^-}-P^\mu_{\pi^-}P^\nu_{\pi^0})\label{ad:3}\,.
\end{align}
\end{subequations}
In the previous equations, the momentum of the meson system is $q^\mu=(P_{\pi^-}+P_{\pi^0})^\mu$, with $s=q^2$. We also introduced $Q^\mu=(P_{\pi^-}-P_{\pi^0})^\mu+(\Delta_{\pi^0\pi^-}/s)q^\mu$, 
and $\Delta_{\pi^0\pi^-}=m_{\pi^0}^2-m_{\pi^-}^2$. Clebsch-Gordan flavor coefficients are $C_S=C_V=\sqrt{2}$ for this decay channel.

The $F_S(s)$ and $F_0(s)$ form factors can be related by taking the divergence of the vector current via 
\begin{equation}
F_S(s)=C_S\frac{\Delta_{\pi^-\pi^0}}{(m_d-m_u)}F_0(s).
\end{equation}
As in ref. \cite{E}, the scalar contribution can be absorbed into the vector current amplitude. This can achieved by replacing 
\begin{equation}
C_S\frac{\Delta_{\pi^-\pi^0}}{s}\longrightarrow C_S\frac{\Delta_{\pi^-\pi^0}}{s}\left[1+\frac{s\,\hat{\epsilon}_S}{m_\tau(m_d-m_u)}\right],
\end{equation}
in eq. (\ref{ad:1}).

Obtaining the $F_0(s)$, $F_+(s)$ and $F_T(s)$ form factors is discussed in the following section.

\section{HADRONIZATION OF THE SCALAR, VECTOR AND TENSOR CURRENTS}\label{sec4}
Lorentz invariance, together with the discrete symmetries of the strong interactions, determine eqs. (\ref{ad:2}) to (\ref{ad:3}). QCD dynamics is encoded in these hadron matrix elements, although it is not possible to determine them using the Lagrangian of the underlying theory unambiguously. Nevertheless, QCD properties are useful in restricting this hadronic input. On the one hand, it is desirable to keep the properties derived from the (very approximate) chiral symmetry of low-energy QCD and from asymptotic strong interactions, where known. On the other, using dispersion relations is ideal to warrant the correct analytic structure of the amplitudes and to comply with unitarity (at least in the elastic region). These properties will be exploited in what follows, as we will briefly review.\\

As shown in ref. \cite{Descotes}, the scalar form factor $F_0(s)$ can be determined in an essentially model-independent way in the low-energy region, though it does not involve resonance contributions to first order in isospin breaking. The S-wave $\pi^-\pi^0$ system must have isospin $I=2$. Watson's final-state interactions theorem \cite{Watson:1952ji} ensures that -in the elastic region- the phase of the di-meson form factor with definite angular momentum ($L$) and isospin ($I$) coincides with the corresponding meson-meson scattering phase shift having the same $L$ and $I$ values ($L=0$ and $I=2$ in our case, so this phase shift is $\delta^2_0(s)$ according to the usual notation). Neglecting inelastic effects (that is a good approximation up to $s\sim 1$ GeV$^2$ in this case), the required di-pion scalar form factor can be obtained \cite{Descotes} by means of a phase dispersive representation 
($F_0(0)=1$ has been used)
\begin{equation}
 F_0(s)\,=\,\mathrm{exp}\left\lbrace \frac{s}{\pi} \int_{4m_\pi^2}^\infty ds' \frac{\delta^2_0(s')}{s'(s'-s-i\epsilon)}\right\rbrace\,,
\end{equation}
since the phase shift $\delta^2_0(s)$ has been measured \cite{Losty:1973et, Hoogland:1977kt}. $|F_0(s)|$ and $\delta^2_0(s)$ are plotted in the upper panel of Fig. 12 in ref. \cite{Descotes}. 
As expected, there is no hint of resonance dynamics in $F_0(s)$.\\

The vector form factor, $F_+(s)$, is known with great accuracy, both theoretically and experimentally. In absence of new-physics interactions, it can be extracted directly from $\tau^-\to\pi^-\pi^0\nu_\tau$ data (since the scalar form factor is negligible up to second-order isospin-violating corrections \cite{Cirigliano:2001er}, which are tiny). The di-pion invariant mass spectrum in these decays has been most precisely measured by the Belle Collaboration \cite{Belle} (it was earlier obtained by the CLEO \cite{Anderson:1999ui}, and 
ALEPH \cite{Schael:2005am} and OPAL \cite{Ackerstaff:1998yj} LEP collaborations). $F_+(s)$ can also be accessed -through a CVC violating correction \cite{Cirigliano:2001er, Cirigliano:2002pv}- via $e^+e^-\to\pi^+\pi^-$ cross-section data at low energies, which has been measured very precisely by BaBar \cite{Lees:2012cj}, BES-III \cite{Ablikim:2015orh}, CMD-2 \cite{Akhmetshin:2006bx}, KLOE-2 \cite{Anastasi:2017eio} and SND \cite{Achasov:2006vp, Akhmetshin:2006wh}. Finally, in the elastic region ($s\lesssim 1$ GeV$^2$), $F_+(s)$ is related via unitarity with the spin-one isospin-one $\pi\pi$ scattering amplitude, for which accurate measurements have been performed \cite{Ochs, Hyams:1973zf, Estabrooks:1974vu}. All previous measurements correspond to the $s>0$ region, $e^-\pi$ scattering \cite{Amendolia:1986wj} probes $F_+(s<0)$.

Theoretically, $F_+(s)$ is well-constrained at low-energies by $\chi PT$ \cite{ChPT1, ChPT2, ChPT3} and in the asymptotic regime by short-distance QCD results \cite{Brodsky:1973kr, Lepage:1980fj}. In the intermediate energy ($\mathcal{O}(1)$ GeV) region, resonance dynamics is needed to interpolate between the two former limits. An adequate tool to connect all energy ranges taking advantage of analyticity and unitarity constraints on $F_+(s)$ are the dispersion relations, which have been employed widely in this context (see i. e. ref. \cite{Dumm} and references therein). We will not discuss at length the procedure here, but only recall that an excellent description of the data can be achieved with three subtractions (one is used to set $F_+(0)=1$)
\begin{equation}
 F_+(s)\,=\,\mathrm{exp}\left[ \alpha_1 s + \frac{\alpha_2}{2}s^2 + \frac{s^3}{\pi} \int_{4m_\pi^2}^\infty ds' \frac{\delta_1^1(s)}{(s')^3(s'-s-i\epsilon)} \right]\,,
\end{equation}
being $\alpha_{1,2}$ the remaining subtraction constants, to be fitted to low-energy data, and $\delta_1^1(s)$ the relevant phase shift. In ref.~\cite{Dumm}, $\delta_1^1(s)$ is given (below the $\rho'$ resonance region), in terms of the $\rho(770)$ pole position and the pion decay constant, $F_\pi$. Its description in the $[M_{\rho'}\lesssim \sqrt{s} \leq M_\tau]$ interval depends on the $\rho'$ and $\rho''$ properties. We will use this framework in what follows. The central values of the modulus and phase of $F_+(s)$ are plotted and compared to data in Figs. 1 and 2 in ref.~\cite{Dumm}. We will use the best fit results corresponding to case III in this reference, which includes first-order isospin breaking corrections. Both statistical and systematic uncertainties on $F_+(s)$ are taking into account throughout our numerical analysis.\\

Although it is difficult to constrain the hadronization of the tensor current, eq. (\ref{ad:3}), from first principles, this would be desirable as it turns out that the $\tau^-\to\pi^-\pi^0\nu_\tau$ decays have the potential to set competitive bounds on (non-standard) charged current tensor interactions. This is in contrast with the $\tau^-\to\pi^-\eta^{(\prime)}\nu_\tau$ decays explored in ref. \cite{E}, which are competitive for new scalar contributions but not for tensor ones, which justified using leading-order $\chi PT$ results for eq. (\ref{ad:3}) in that analysis. Unfortunately, there is no experimental data that can guide us in building $F_T(s)$, so will rely only on theory to accomplish this task.

Since $s$ can vary from the two-pion threshold up to $M_\tau^2$, light resonances contribution (giving the energy dependence of the form factor) should be included in a refined analysis, as we intend. We show in the appendix that, for $F_T(s)$, it is not convenient to extend the energy range of applicability of $\chi PT$ by including the resonances as explicit degrees of freedom, in the so-called Resonance Chiral Theory \cite{Ecker}. Instead, it will be more appropriate to use a dispersive construction of $F_T(s)$ taking advantage of unitarity constraints on its phase \cite{Cirigliano:2017tqn}. $F_T(0)$ will be studied within $\chi PT$ in the following.

The lowest-order $\chi PT$ Lagrangian with tensor sources, which is $\mathcal{O}(p^4)$ in the chiral counting \cite{Mateu}, includes only four operators. Among them, only the one with coefficient $\Lambda_2$ contributes to the studied decays:
\begin{equation}\label{ad:7}
\mathcal{L}=\Lambda_1 \langle t^{\mu\nu}_+ f_{+\mu\nu}\rangle -i\Lambda_2 \langle t^{\mu\nu}_+ u_\mu u_\nu \rangle + \dots.
\end{equation}
In the preceding equation, $t^{\mu\nu}_+ =u^{\dagger}t^{\mu\nu}u^{\dagger}+u t^{\mu\nu\dagger}u$ and $\left\langle\cdots\right\rangle$ means a flavor space trace. Operators in eq. (\ref{ad:7}) are built with chiral tensors \cite{Bijnens:1999sh}, with three of them entering the displayed operators:
\begin{itemize}
\item $u_\mu=i\left[u^\dagger (\partial_\mu-i r_\mu)u-u(\partial_\mu -i l_\mu)u^\dagger \right]$, which includes the left- and right-handed sources, $\ell_\mu$ and $r_\mu$.
\item The chiral tensor sources $t^{\mu\nu}$ and its adjoint, and
\item $f_+^{\mu\nu}=u F_L^{\mu\nu}u^{\dagger}+u^{\dagger}F_R^{\mu\nu}u$, including the left- and right-handed field-strength tensors, $F_L^{\mu\nu}$ and $F_R^{\mu\nu}$, given in terms of $\ell^\mu$ and $r^\mu$.
\end{itemize}

Let us recall the non-linear representation of the pseudo Goldstone bosons, given by $u=\exp \left[\frac{i}{\sqrt{2}F}\phi\right]$ \cite{Coleman:1969sm, Callan:1969sn}, where (for two flavors)
\begin{equation}\label{ad:matrix}
\phi=\left(\begin{array}{cc}
\frac{\pi^0}{\sqrt{2}} & \pi^+\\
\pi^- & -\frac{\pi^0}{\sqrt{2}}
\end{array}\right),
\end{equation}
$F$ being the pion decay constant in the chiral limit, $F\sim F_\pi \sim 92$ MeV. All resonance multiplets considered below have analogous flavor structure to eq.~(\ref{ad:matrix}). 

The tensor source ($\bar{t}^{\mu\nu}$) is related to its chiral projections ($t^{\mu\nu}$ and $t^{\mu\nu\dagger}$) by means of \cite{Mateu} 
\begin{equation}
t^{\mu\nu}=P_L^{\mu\nu\lambda\rho}\bar{t}_{\lambda\rho},\qquad 4P_L^{\mu\nu\lambda\rho}=(g^{\mu\lambda}g^{\nu\rho}-g^{\mu\rho}g^{\nu\lambda}+i\epsilon^{\mu\nu\lambda\rho}),
\end{equation}
where $\bar{\Psi}\sigma_{\mu\nu}\bar{t}^{\mu\nu}\Psi$ is the tensor quark current.

From eq. (\ref{ad:7}) it can be shown \cite{E} that, in the limit of isospin symmetry~\footnote{Since $F_T(s)$, as given by eq.~(\ref{T m.e.}), is purely real and the sign of $\Lambda_2$ was unknown, a factor $i$ was absorbed redefining $F_T(s)$ in ref.~\cite{E}. As we consider a non-vanishing tensor form factor phase (see eq.(\ref{DRFT}) and related discussion), we will not follow this procedure in the present analysis.},
\begin{equation}
i\left\langle \pi^-\pi^0\left\vert \frac{\delta \mathcal{L}^{\mathcal{O}(p^4)}_{\chi PT}}{\delta \bar{t}_{\alpha\beta}}\right\vert 0\right\rangle =\frac{\sqrt{2}\Lambda_2}{F^2}\left(p_{\pi^-}^\alpha p_{\pi^0}^\beta-p_{\pi^0}^\alpha p_{\pi^-}^\beta \right).
\label{T m.e.}
\end{equation}

We show in the appendix that it is not convenient to include the energy-dependence of the tensor form factor by extending $\chi PT$ \cite{ChPT1, ChPT2, ChPT3} including resonances \cite{Ecker, Ecker:1989yg}.

Ref. \cite{Baum:2011rm} evaluated $f_T(0)=2m_\pi F_T(0)$ on the lattice. Their result, $f_T(0)=0.195\pm0.010$ yields $\Lambda_2\,=\,(12.0\pm0.6)$ MeV, that we will use in the following. This value of $\Lambda_2$ is roughly a factor three smaller than the prediction for $\Lambda_1$ obtained using short-distance QCD properties \cite{Mateu:2007tr}, $\Lambda_1\,=\,(33\pm2)$ MeV. Since both operators displayed in eq.~(\ref{ad:7}) have the same chiral counting order, one would have guessed $\Lambda_2\sim\Lambda_1$, resulting in an overestimation of $\Lambda_2$, as we did in ref.~\cite{E}~\footnote{Fortunately, since the $\tau^-\to \eta^{(\prime)} \pi^-\nu_\tau$ decays are quite insensitive to tensor interactions, this does not change the limits obtained in this paper for $\hat{\epsilon}_S$.}.

\begin{figure}[t]
	\includegraphics[width=7cm]{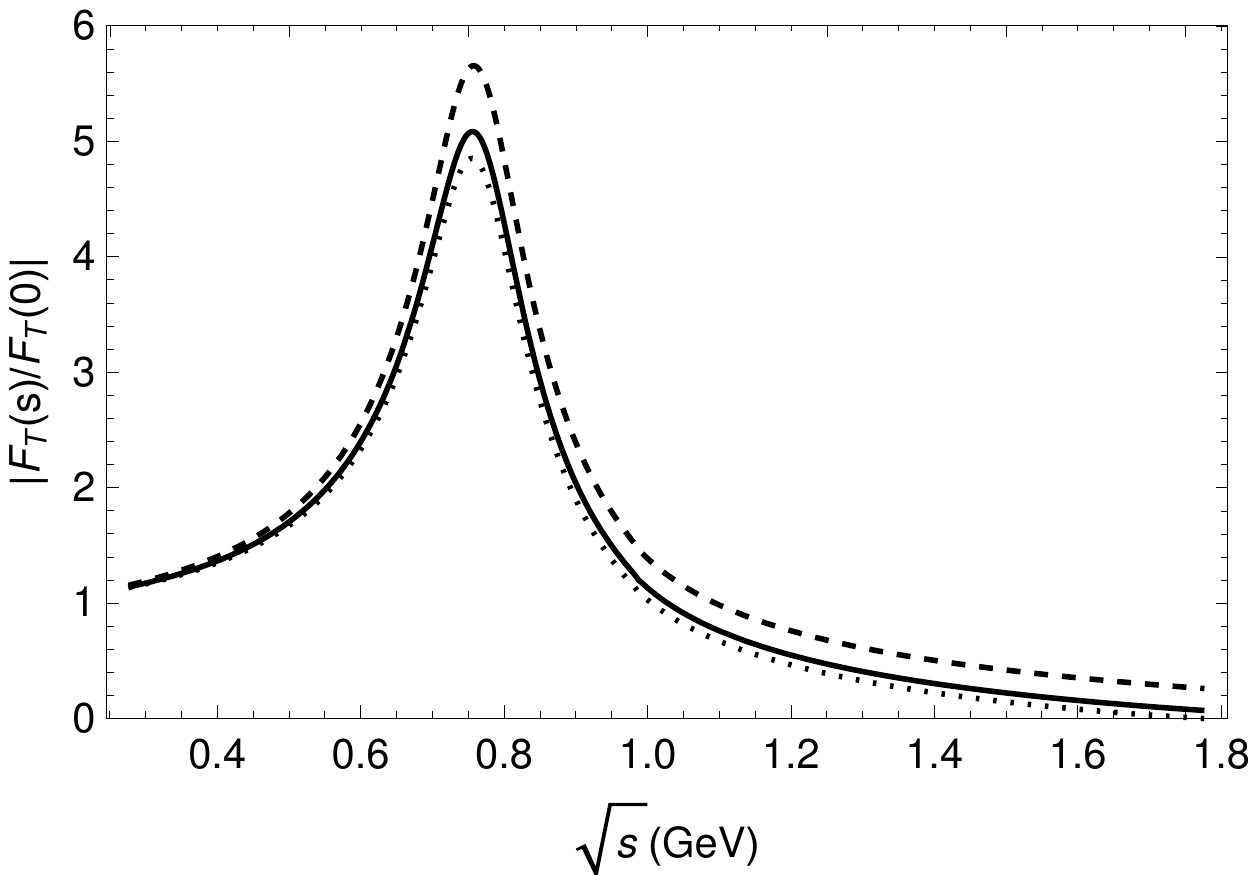}
	\includegraphics[width=7cm]{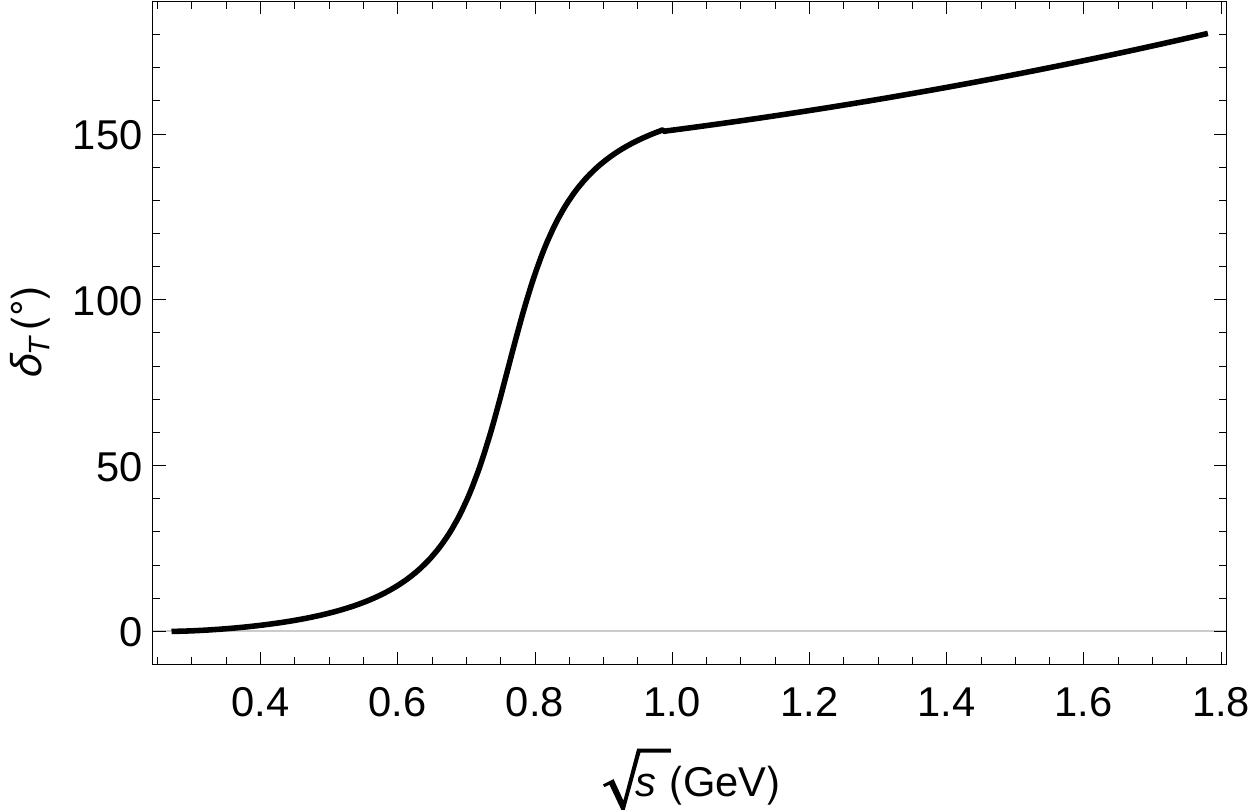}
	\centering
	\caption{Modulus and phase, $|F_T(s)|$ (left) and $\delta_T(s)$ (right), of the tensor form factor, $F_T(s)$, corresponding to eq. (\ref{DRFT}).}\label{PlotsFTsfinal}
\end{figure}

We will follow ref. \cite{Cirigliano:2017tqn} and obtain $F_T(s)$ using again a phase dispersive representation. As shown in ref. \cite{Cirigliano:2017tqn} (see also the appendix of this article), the tensor form factor phase equals the vector form factor phase, $\delta_T(s)=\delta_+(s)$, in the elastic region. We will use the previous equation also above the onset of inelasticities in our dispersion relation 
\begin{equation}
\frac{F_T(s)}{F_T(0)}\,=\,\mathrm{exp}\left\lbrace \frac{s}{\pi} \int_{4m_\pi^2}^\infty ds' \frac{\delta_T(s')}{s'(s'-s-i\epsilon)}\right\rbrace\,,\label{DRFT}
\end{equation}
and fix $F_T(0)=\frac{\sqrt{2}\Lambda_2}{F^2}$ according to the leading-order $\chi PT$ result. We plot in figure \ref{PlotsFTsfinal} the modulus and phase of $F_T(s)$ obtained using eq. (\ref{DRFT}). The different curves on the left panel are obtained for $s_{max}=M_\tau^2$, $4$ and $9$ GeV$^2$ \footnote{The parameter $s_{max}$ corresponds to the cutoff of the dispersive integral. The unphysical dependence on it is a consequence of the dispersion relation (\ref{DRFT}) being once-subtracted. Additional subtractions would reduce the artificial dependence on $s_{max}$. However, since we lack low-energy information to fix these subtraction constants, we cannot follow this procedure. Taking this into account, we restrict the $s_{max}$ values in the previously quoted range.} and we will take this range for $F_T(s)$ as an estimate of our corresponding error (our plots will be given for $s_{max}=4$ GeV$^2$ in the following). We neglect the uncertainty associated to our ignorance on the inelasticities affecting $\delta_T(s)$ (see the related discussion in ref. \cite{Cirigliano:2017tqn}), which are small below  $\sqrt{s}=1.3$ GeV.

\section{DECAY OBSERVABLES}\label{sec5}
In order to study possible NP effects in these decays, one should use not only the hadronic spectrum and branching ratio, but also  Dalitz plot distributions and the measurable forward-backward asymmetry. In this section, we focus in the study of the possible effects of the non-standard effective couplings described in section \ref{sec2} in these $\tau^-\to\pi^-\pi^0\nu_\tau$ decay observables. We will start with the Dalitz plots (which should contain more dynamical information, as no integration over any of the two independent kinematical variables has been performed) and move later on to (partially) integrated observables: differential decay rate as function of the di-meson invariant mass, forward-backward asymmetry and, finally, branching ratio.

The differential decay width of the $\tau^-\to\pi^-\pi^0\nu_\tau$ decays, in the $\tau$ lepton rest frame, is
\begin{equation}\label{DO:eq1}
\frac{d^2\Gamma}{ds dt}=\frac{1}{32(2\pi)^3M_\tau^3}\overline{\vert\mathcal{M}\vert^2},
\end{equation} 
where $\overline{\vert\mathcal{M}\vert^2}$ represents the unpolarized spin-averaged squared matrix element, $s$ being the $\pi^0\pi^-$ system invariant mass, limited in the interval  $(m_{\pi^0}+m_{\pi^-})^2\leq s \leq M_\tau^2$ and $t=(p'+p_{\pi^0})^2=(p-p_{\pi^-})^2$ with $t^-(s)\leq t \leq t^+(s)$, where
\begin{equation}
t^\pm(s)=\frac{1}{2s}\left[2s(M_\tau^2+m_{\pi^0}^2-s)-(M_\tau^2-s)(s+m_{\pi^-}^2-m_{\pi^0})\pm (M_\tau^2-s)\sqrt{\lambda(s,m_{\pi^-}^2,m_{\pi^0}^2)}\right],
\end{equation}
and $\lambda(x,y,z)=x^2+y^2+z^2-2xy-2xz-2yz$ is the usual Kallen function.

\subsection{Dalitz plot}
Including possible non-standard weak charged current interactions, the unpolarized spin-averaged squared amplitude  yields \footnote{We note a typo writing the corresponding equation, (22), of ref.~\cite{E}, where the factor $2$ should not appear. All subsequent expressions and the numerical results of ref. \cite{E} are not affected by this typo.}
\begin{equation}\label{Dp:eq5}
\overline{\vert\mathcal{M}\vert^2}=\frac{G_F^2 |V_{ud}|^2S_{EW}}{s^2}\left(1+\epsilon_L+\epsilon_R\right)^2\left[M_{00}+M_{++}+M_{0+}+M_{T+}+M_{T0}+M_{TT}\right],
\end{equation}
where the scalar, vector and tensor squared amplitudes are $M_{00}$, $M_{++}$ and $M_{TT}$, respectively. Their corresponding interferences are denoted $M_{0+}$, $M_{T+}$, $M_{T0}$. All these read \footnote{Comparing eqs. (\ref{ad:2}) to (\ref{ad:3}) to their analogs in ref. \cite{E}, it can be verified that eqs.  (\ref{Dpcontributions}) agree with the corresponding expressions in ref. \cite{E}.}
\begin{equation}\begin{split}
M_{0+}&=2C_V \,C_S \,m^2_\tau \,\mathrm{Re}\left[F_+(s)F_0^*(s)\right]\,\Delta_{\pi^-\pi^0}\left(1+\frac{s\hat{\epsilon}_S}{m_\tau(m_d-m_u)}\right)\\
&\qquad\times \left\lbrace s\left(m_\tau^2-s-2t+\Sigma_{\pi^-\pi^0}\right)-m^2_\tau\Delta_{\pi^-\pi^0}\right\rbrace,\\
M_{T+}&=4C_V\,\hat{\epsilon}_T\,m_\tau^3\, s\, \mathrm{Re}\left[F_T(s)F_+^*(s)\right]\left(1-\frac{s}{m_\tau^2}\right)\lambda(s,m_{\pi^-}^2,m_{\pi^0}^2),\\
M_{T0}&=4C_S\,\Delta_{\pi^-\pi^0}\,\hat{\epsilon}_T\, m_\tau\, s\,\mathrm{Re}\left[F_T(s)F_0^*(s)\right]\left(1+\frac{s\hat{\epsilon}_S}{m_\tau(m_d-m_u)}\right)\\
&\qquad\times \left\lbrace s\left(m^2_\tau-s-2t+\Sigma_{\pi^-\pi^0}\right)-m^2_\tau\Delta_{\pi^-\pi^0}\right\rbrace,\\
M_{00}&=C_S^2\,\left(\Delta_{\pi^-\pi^0}\right)^2m_\tau^4\left(1-\frac{s}{m_\tau^2}\right)|F_0(s)|^2\left(1+\frac{s\hat{\epsilon}_S}{m_\tau(m_d-m_u)}\right)^2,\\
M_{++}&=C_V^2\,|F_+(s)|^2\biggl\lbrace m^4_\tau \left(s-\Delta_{\pi^-\pi^0}\right)^2-m_\tau^2 s \left[s(s+4t)-2\Delta_{\pi^-\pi^0}\left(s+2t-\Sigma_{\pi^-\pi^0}\right)+\left(\Delta_{\pi^-\pi^0}\right)^2\right]\\
&\qquad\qquad+4m_{\pi^-}^2s^2\left(m^2_{\pi^0}-t\right)+4s^2t\left(s+t-m^2_{\pi^0}\right)\biggr\rbrace,\\
M_{TT}&=4\hat{\epsilon}_T^2 \,|F_T(s)|^2s^2\biggl\lbrace m_{\pi^-}^4\left(m_\tau^2-s\right)-2m_{\pi^-}^2\left(m_\tau^2-s\right)\left(s+2t-m_{\pi^0}^2\right)-m_{\pi^0}^4\left(3m_\tau^2+s\right)\\
&\qquad\qquad+2m_{\pi^0}^2\left[\left(s+m_\tau^2\right)\left(s+2t\right)-2m_\tau^4\right]-s\left[\left(s+2t\right)^2-m_\tau^2\left(s+4t\right)\right]\biggr\rbrace ,\label{Dpcontributions}
\end{split}\end{equation}
where the familiar definitions $\Delta_{\pi^-\pi^0}=m_{\pi^-}^2-m_{\pi^0}^2$ and $\Sigma_{\pi^-\pi^0}=m_{\pi^-}^2+m_{\pi^0}^2$ were employed. Noteworthy, the scalar form factor is always suppressed by $\Delta_{\pi^-\pi^0}$, which is tiny, in the previous equations for $M_{00}$, $M_{T0}$ and $M_{0+}$. This makes its effect negligible even for $|\hat{\epsilon}_S|\sim 1$ (radiative pion decay limits $|\hat{\epsilon}_S|\lesssim 0.01$ and, under the reasonable assumption of lepton flavor universality, this limit should also apply for the tau flavor considered here).

\begin{figure}[t]
\includegraphics[width=7cm]{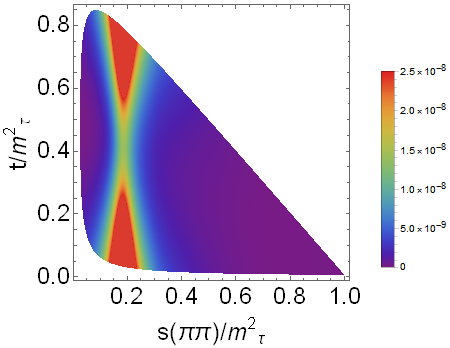}	
\includegraphics[width=7cm]{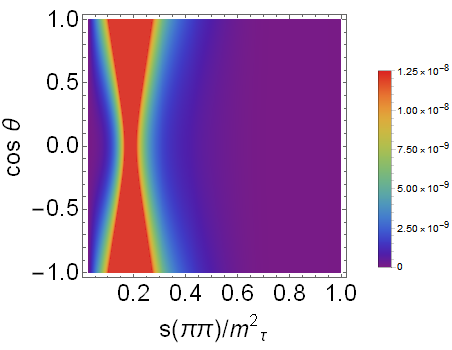}		
\centering			
\caption{Dalitz plot distribution $\overline{\vert \mathcal{M}\vert^2}_{00}$ in the SM, eq. (\ref{Dp:eq5}): Differential decay distribution for $\tau^-\to\pi^-\pi^0\nu_\tau$ in the (s,t) variables (left). The right-hand figure shows the differential decay distribution in the $(s,\cos \theta)$ variables, eq. (\ref{a2:eq4}). The Mandelstam variables, s and t, are normalized to $M_\tau^2$.}\label{Obs:fig1}
\end{figure}

We now turn to analyze possible NP signatures in Dalitz plots distributions. The left panel of figure \ref{Obs:fig1} shows the squared matrix element $\overline{\vert \mathcal{M}\vert^2}_{00}$ in the (s,t) plane, which is obtained using the SM predictions for $\tau^-\to\pi^-\pi^0\nu_\tau$ form factors \cite{Dumm, Descotes}. The $\rho(770)$ meson dominance of the dynamics is clearly seen in this plot.

In order to better appreciate the modifications induced by non-vanishing $\hat{\epsilon}_{S,T}$ in Dalitz plots, we introduce the observable
\begin{equation}\label{DeltaDalitz}
 \tilde{\Delta}(\hat{\epsilon}_S,\hat{\epsilon}_T)\,=\,\frac{\Bigg|\overline{|\mathcal{M}(\hat{\epsilon}_S,\hat{\epsilon}_T)|^2}-\overline{|\mathcal{M}(0,0)|^2}\Bigg|}{\overline{|\mathcal{M}(0,0)|^2}}\,.
\end{equation}

\begin{figure}[t]
\includegraphics[width=7cm]{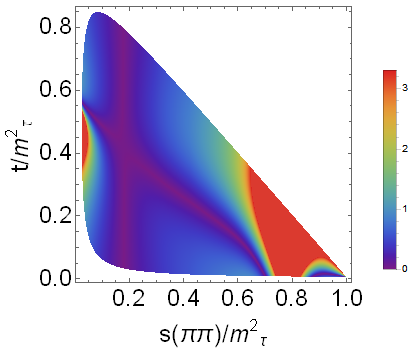}
\includegraphics[width=7cm]{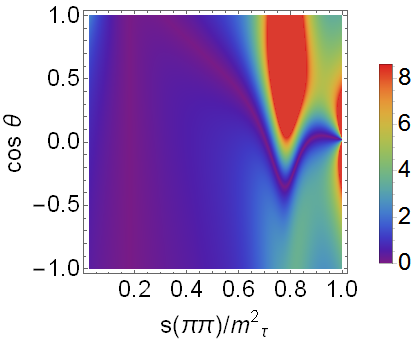}	
\centering			
\caption{Dalitz plot distribution for $\tilde{\Delta}(\hat{\epsilon}_S,\hat{\epsilon}_T)$, (\ref{DeltaDalitz}), in the $\tau^-\to\pi^-\pi^0\nu_\tau$ decays: left-hand side corresponds to eq. (\ref{Dp:eq5}) and right-hand side corresponds to the differential decay distribution in the $(s,\cos \theta)$ variables, both with $(\hat{\epsilon}_S=1.31,\hat{\epsilon}_T=0)$. The Mandelstam variables, s and t, are normalized to $M_\tau^2$.}\label{Obs:fig2}
\end{figure}

\begin{figure}[t]
\includegraphics[width=7cm]{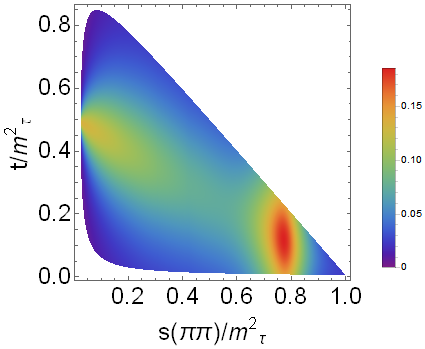}
\includegraphics[width=7cm]{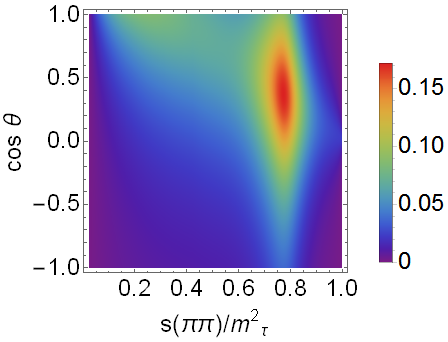}	
\centering			
\caption{Dalitz plot distribution for $\tilde{\Delta}(\hat{\epsilon}_S,\hat{\epsilon}_T)$, (\ref{DeltaDalitz}), in the $\tau^-\to\pi^-\pi^0\nu_\tau$ decays: left-hand side corresponds to eq. (\ref{Dp:eq5}) and right-hand side corresponds to the differential decay distribution in the $(s,\cos \theta)$ variables, both with $(\hat{\epsilon}_S=0,\hat{\epsilon}_T=-0.014)$. The Mandelstam variables, s and t, are normalized to $M_\tau^2$.}\label{Obs:fig3}
\end{figure}

\begin{figure}[t]
\includegraphics[width=7cm]{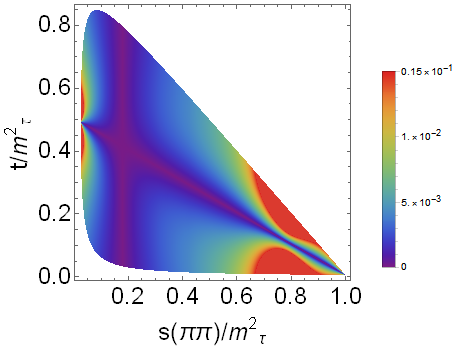}
\includegraphics[width=7cm]{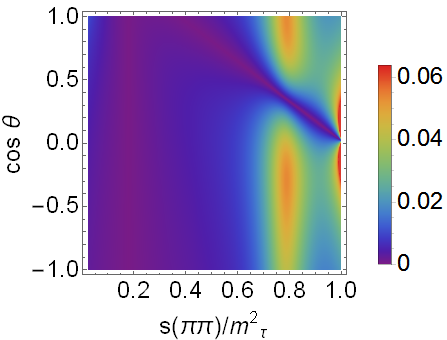}	
\centering			
\caption{Dalitz plot distribution for $\tilde{\Delta}(\hat{\epsilon}_S,\hat{\epsilon}_T)$, (\ref{DeltaDalitz}), in the $\tau^-\to\pi^-\pi^0\nu_\tau$ decays: left-hand side corresponds to eq. (\ref{Dp:eq5}) and right-hand side corresponds to the differential decay distribution in the $(s,\cos \theta)$ variables, both with $(\hat{\epsilon}_S=0.008, \hat{\epsilon}_T=0)$. The Mandelstam variables, s and t, are normalized to $M_\tau^2$.}\label{Obs:fig4}
\end{figure}

\begin{figure}[t]
\includegraphics[width=7cm]{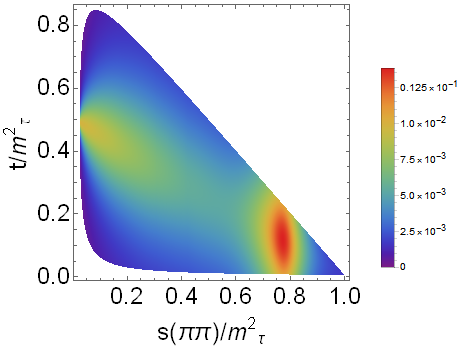}	
\includegraphics[width=7cm]{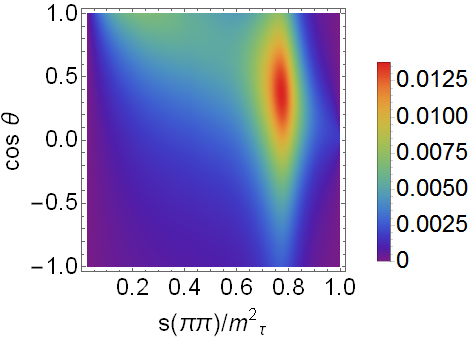}	
\centering			
\caption{Dalitz plot distribution for $\tilde{\Delta}(\hat{\epsilon}_S,\hat{\epsilon}_T)$, (\ref{DeltaDalitz}), in the $\tau^-\to\pi^-\pi^0\nu_\tau$ decays: left-hand side corresponds to eq. (\ref{Dp:eq5}) and right-hand side corresponds to the differential decay distribution in the $(s,\cos \theta)$ variables, both with $(\hat{\epsilon}_S=0,\hat{\epsilon}_T=-0.001)$. The Mandelstam variables, s and t, are normalized to $M_\tau^2$.}\label{Obs:fig5}
\end{figure}

In the left panel of figures \ref{Obs:fig2} and \ref{Obs:fig3}, $\tilde{\Delta}(\hat{\epsilon}_S,\hat{\epsilon}_T)$ (\ref{DeltaDalitz}) is shown for two representative values of the set of 
$(\hat{\epsilon}_S,\hat{\epsilon}_T)$ parameters that are consistent with the $\mathrm{BR}(\tau^-\to\pi^-\pi^0\nu_\tau)$ (obtaining these limits will be discussed in subsection \ref{limits}). Although $\mathcal{O}(1)$ effects are seen in fig. \ref{Obs:fig2}, these are not realistic since two-pion tau decays are almost insensitive to $\hat{\epsilon}_S$. Indeed, when $\hat{\epsilon}_S$ is taken from more adequate processes \cite{a24,a25,E, Cirigliano:2018dyk}, the left panel of fig. \ref{Obs:fig4} shows that only a measurement of $\tilde{\Delta}$ with $\lesssim1\%$ uncertainty could distinguish these new physics effects. In the left plot of fig. \ref{Obs:fig3} (with $(\hat{\epsilon}_S=0,\hat{\epsilon}_T=-0.014)$) the deviations with respect to the SM are around $15\%$ in a given region, but the left plot in figure \ref{Obs:fig5} (obtained using our best fit value for $\hat{\epsilon}_T$ in section \ref{limits}) reduces the size of this signal to a $1\%$ effect. These $\mathcal{O}(1\%)$ effects would be difficult to measure, even at Belle-II \cite{Kou:2018nap}. Our uncertainties do not affect the conclusions drawn in this paragraph.

\subsection{Angular distribution}
The hadronic mass and angular distributions are also modified by the generic new effective interactions that we are studying and can have different sensitivity to $\hat{\epsilon}_S$ and $\hat{\epsilon}_T$. The rest frame of the hadronic system is convenient for this analysis. It is defined by $\vec{p}_{\pi^-}+\vec{p}_{\pi^0}=\vec{p}_\tau-\vec{p}_\nu=0$. In this frame, the charged particle energies are given by $E_\tau=(s+M_\tau^2)/2\sqrt{s}$ and $E_{\pi^-}=(s+m_{\pi^-}^2-m_{\pi^0}^2)/2\sqrt{s}$. The measurable angle $\theta$ between these two particles can be obtained from the invariant $t$ variable by means of $t=m_{\pi^-}^2+m_\tau^2-2E_\tau E_{\pi^-}+2\vert \vec{p}_{\pi^-}\vert \vert \vec{p}_\tau\vert \cos \theta$, with $\vert\vec{p}_{a}\vert=\sqrt{E^2_a-m^2_a}$ for $a=\pi^-,\tau^-$.

The Dalitz decay distribution in the $(s,\cos \theta)$ variables, for general $\hat{\epsilon}_S$ and $\hat{\epsilon}_T$ reads
\begin{equation}\label{a2:eq4}\begin{split}
\frac{d^2\Gamma}{d\sqrt{s}d\cos\theta}&=\frac{G_F^2|V_{ud}|^2S_{EW}}{128\pi^3 m_\tau}(1+\epsilon_L+\epsilon_R)^2\left(\frac{m_\tau^2}{s}-1\right)^2|\vec{p}_{\pi^-}|\biggl\lbrace C_S^2\left(\Delta_{\pi^- \pi^0}\right)^2|F_0(s)|^2\\
&\quad \times \left( 1+\frac{s\hat{\epsilon}_S}{m_\tau (m_d-m_u)}\right)^2+16|\vec{p}_{\pi^-}|^2s^2\left\vert\frac{C_V}{2m_\tau}F_+(s)+\hat{\epsilon}_T F_T(s)\right\vert^2\\
&+4|\vec{p}_{\pi^-}|^2s\left(1-\frac{s}{m_\tau^2}\right)\cos^2\theta\left[C_V^2|F_+(s)|^2-4s\hat{\epsilon}_T^2|F_T(s)|^2\right]-4C_S\Delta_{\pi^-\pi^0}|\vec{p}_{\pi^-}|\sqrt{s}\cos\theta \\
&\quad\times\left( 1+\frac{s\hat{\epsilon}_S}{m_\tau (m_d-m_u)}\right)\left[C_V \mathrm{Re}\left[F_0(s)F_+^*(s)\right]+\frac{2s \hat{\epsilon}_T}{m_\tau}\mathrm{Re}\left[F_T(s)F_0^*(s)\right]\right]\biggr\rbrace,
\end{split}\end{equation}
which coincides with the SM result when these two effective NP couplings are set to zero.

The right panel of figure \ref{Obs:fig1} shows eq. (\ref{a2:eq4}) for $\pi^-\pi^0$ in the SM case. In the right panel of figures \ref{Obs:fig2} and \ref{Obs:fig3} the ($s,\,\cos \theta$) distributions for $\tilde{\Delta}(\hat{\epsilon}_S,\hat{\epsilon}_T)$, (\ref{DeltaDalitz}), are plotted; for the same representative values of ($\hat{\epsilon}_S$, $\hat{\epsilon}_T$) used in order to obtain the left panel of these figures. Again for non-standard scalar interactions, the large effect seen in the left panel of fig. \ref{Obs:fig2} is unrealistic and it will be  challenging to measure the reduced effect ($\lesssim 6\%$) of fig. \ref{Obs:fig4} at Belle-II \cite{Kou:2018nap}. For tensor interactions, the deviation from the SM depicted in the right plot of fig. \ref{Obs:fig3} could be measurable, but this is not the case for the effect seen in the right plot of figure \ref{Obs:fig5} ($\lesssim 1\%$), obtained using our preferred value for $\hat{\epsilon}_T$. Again, our uncertainties do not affect the preceding discussion.

\subsection{Decay rate}
\begin{figure}[h!]
\includegraphics[width=12cm]{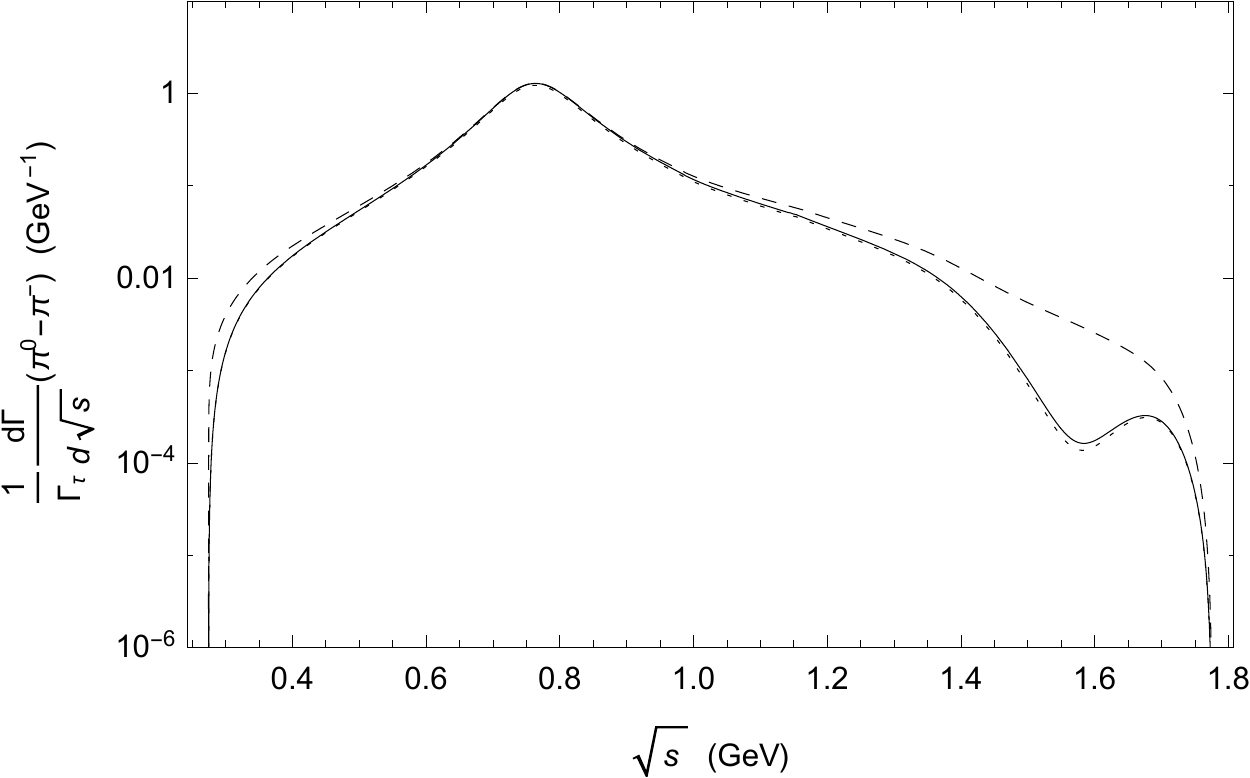}		
\centering			
\caption{The $\pi^0\pi^-$ hadronic invariant mass distribution for the SM (solid line) and $\hat{\epsilon}_S=1.31,\hat{\epsilon}_T=0$ (dashed line), $\hat{\epsilon}_S=0,\hat{\epsilon}_T=-0.014$ (dotted line). Axes units are given in $\mathrm{GeV}$ powers and the decay distributions are normalized to the tau decay width.}\label{Obs:fig6}
\end{figure}

The di-pion invariant mass distributions is obtained integrating upon the $t$ variable in eq. (\ref{DO:eq1})
\begin{equation}\begin{split}\label{Dr:eq1}
\frac{d\Gamma}{d s}=&\frac{G_F^2\vert V_{ud}\vert^2 m_\tau^3 S_{EW}}{384\pi^3 s}\left(1+\epsilon_L+\epsilon_R\right)^2\left(1-\frac{s}{m_\tau^2}\right)^2\lambda^{1/2}\left(s,m_{\pi^0}^2,m_{\pi^-}^2\right)\\
& \qquad\times\left[X_{VA}+\hat{\epsilon}_SX_S+\hat{\epsilon}_TX_T+\hat{\epsilon}_S^2X_{S^2}+\hat{\epsilon}_T^2X_{T^2}\right],
\end{split}\end{equation}
where
\begin{subequations}
\begin{align}
&X_{VA}=\frac{1}{2s^2}\left[3\vert F_0(s)\vert^2C_S^2 \Delta_{\pi^-\pi^0}^2+\vert F_+(s)\vert^2C_V^2\left(1+\frac{2s}{m_\tau^2}\right)\lambda\left(s,m_{\pi^0}^2,m_{\pi^-}^2\right)\right],\\
&X_S=\frac{3}{s\, m_\tau}\vert F_0(s)\vert^2C_S^2\frac{\Delta_{\pi^-\pi^0}^2}{m_d-m_u},\\
&X_T=\frac{6}{s\,m_\tau}\mathrm{Re}\left[F_T(s)F_+^*(s)\right]C_V\lambda\left(s,m_{\pi^0}^2,m_{\pi^-}^2\right),\\
&X_{S^2}=\frac{3}{2\,m_\tau^2}\vert F_0(s)\vert^2 C_S^2\frac{\Delta_{\pi^-\pi^0}^2}{\left( m_d-m_u\right)^2},\\
&X_{T^2}=\frac{4}{s}\vert F_T(s)\vert^2 \left(1+\frac{s}{2\,m_\tau^2}\right)\lambda\left(s,m_{\pi^0}^2,m_{\pi^-}^2\right).
\end{align}
\end{subequations}
Again, the SM limit is recovered with $\epsilon_L=\epsilon_R=\hat{\epsilon}_S=\hat{\epsilon}_T=0$. Figure \ref{Obs:fig6} plots the invariant mass distribution of the di-pion system for $\tau^-\to\pi^-\pi^0\nu_\tau$ decays. It is almost impossible to distinguish the case of tensor interactions from the SM curve and, although some departure is seen for non-standard scalar interactions, it goes away when realistic values on $|\hat{\epsilon}_S|\sim10^{-2}$ \cite{a24,a25,E} are considered.

\subsection{Forward-backward asymmetry}
\begin{figure}[h!]
			\includegraphics[width=7.5cm]{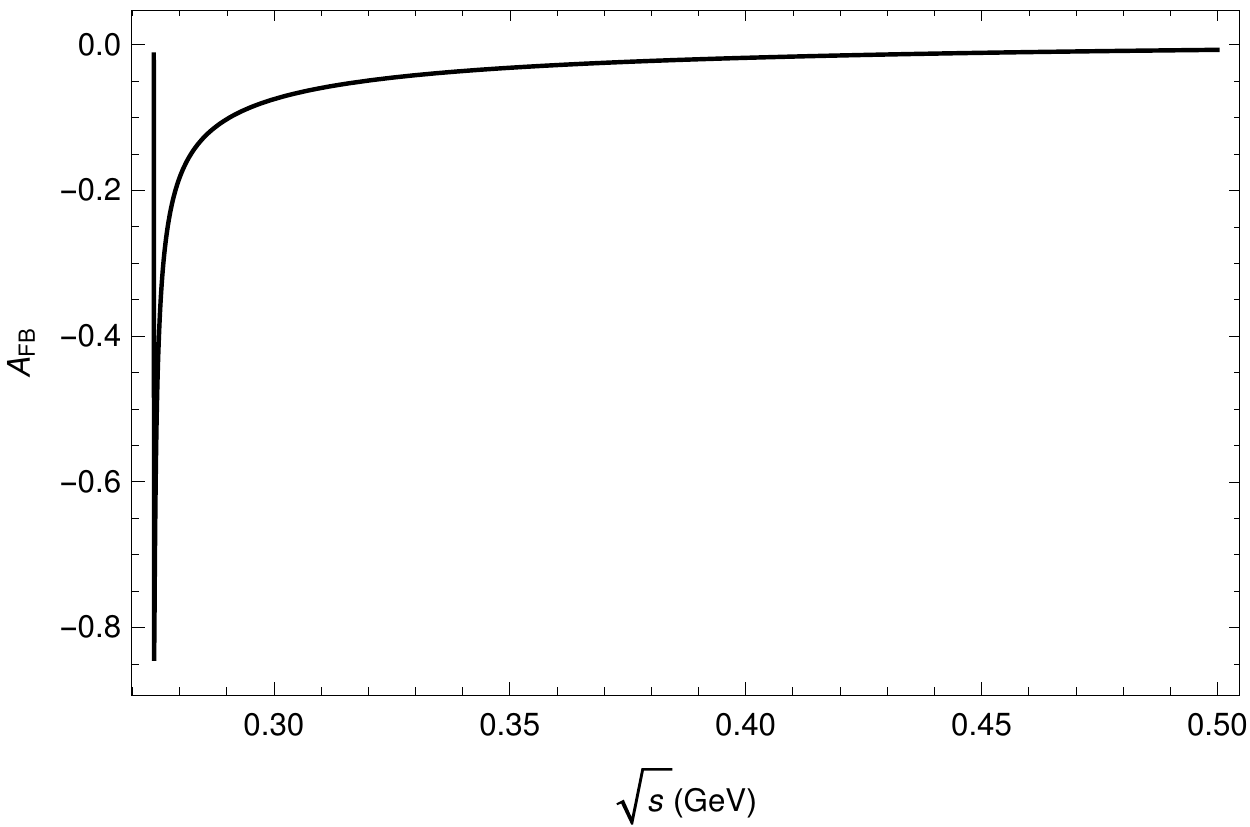}
			\includegraphics[width=7.5cm]{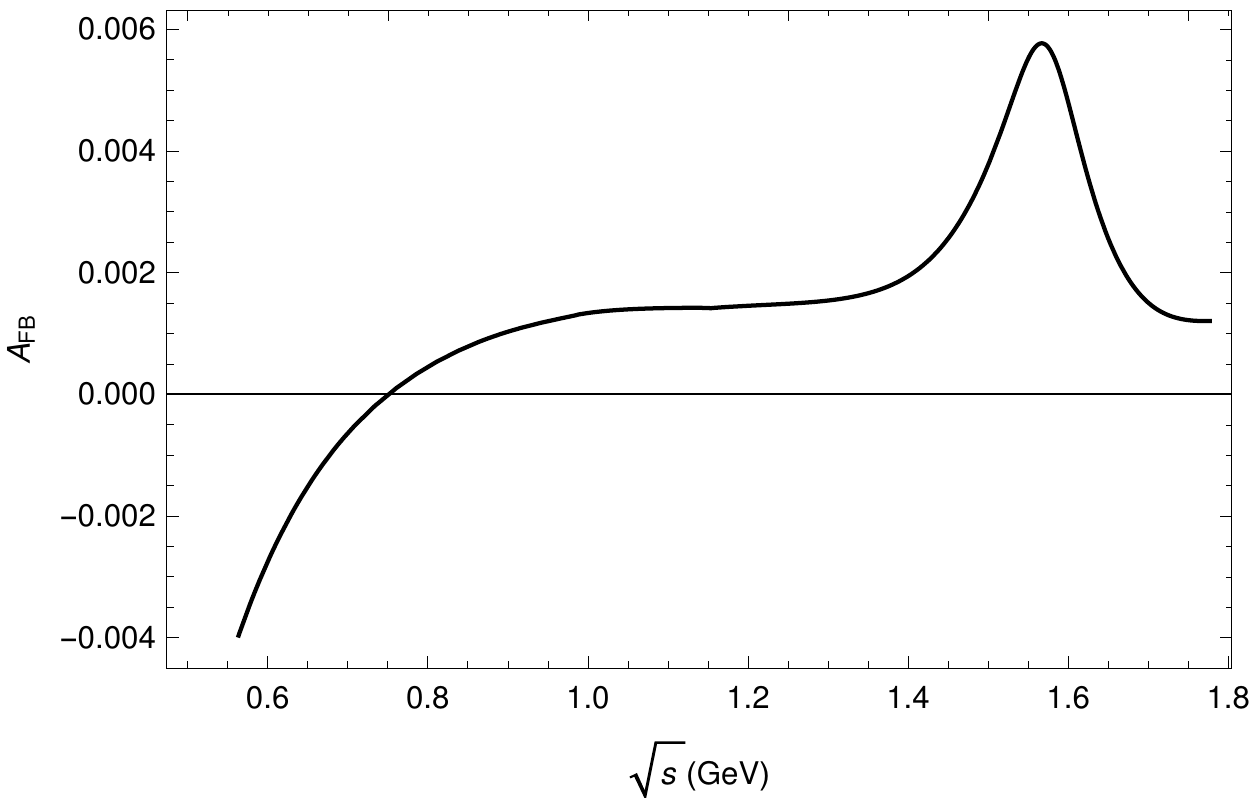}	
			\centering
			\caption{The forward-backward asymmetry in the $\tau^-\to\pi^-\pi^0\nu_\tau$ decay as a function of the $\pi\pi$ energy for the SM case. The low-energy region is shown in the left plot and remaining energy range is represented in the right plot.}
			\label{As:fig}
\end{figure}

\begin{figure}[h!]
\includegraphics[width=7cm]{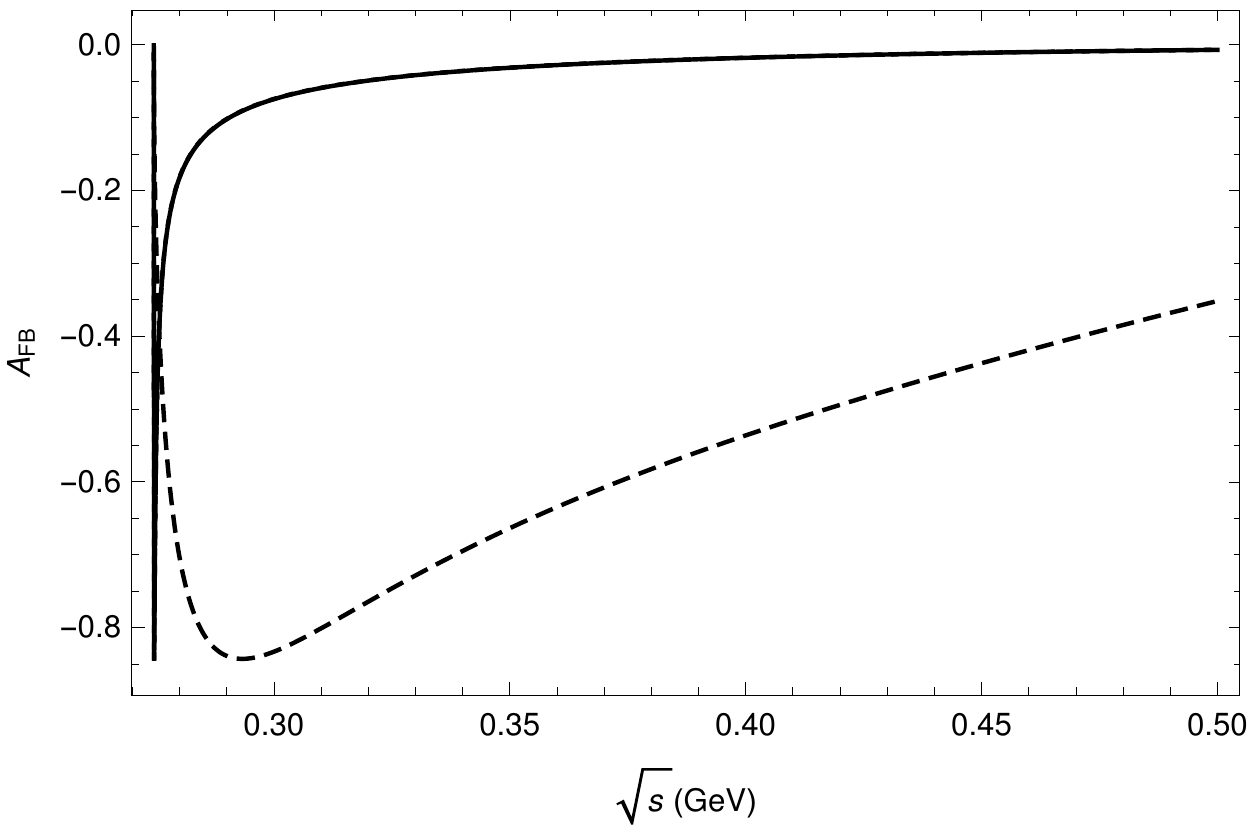}
\includegraphics[width=7cm]{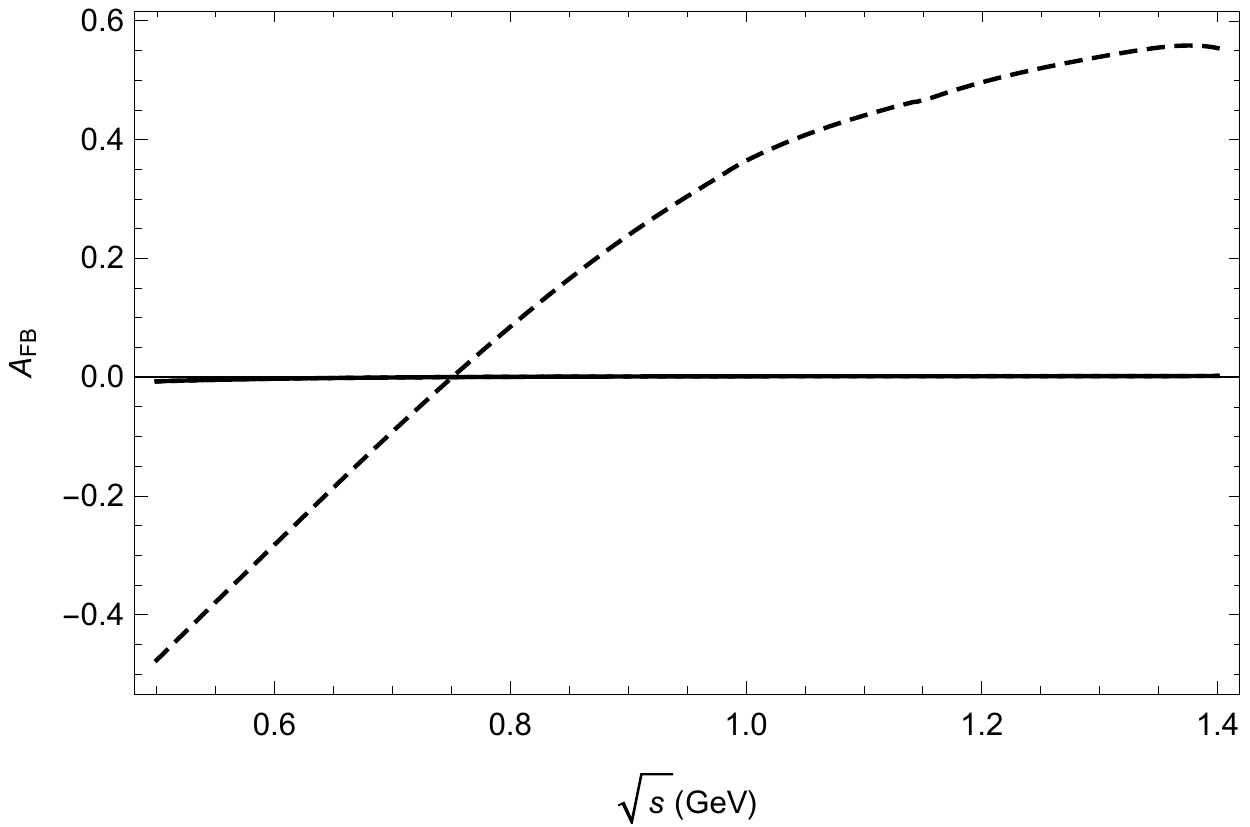}	
\centering			
\caption{Forward-asymmetry for $\hat{\epsilon}_S=1.31,\hat{\epsilon}_T=0$ (dashed line) compared to the SM prediction (solid line). The left plot shows the low-energy region and the right plot includes the remaining energy range.}\label{Obs:fig7} 
\end{figure}

\begin{figure}[h!]
	\includegraphics[width=7cm]{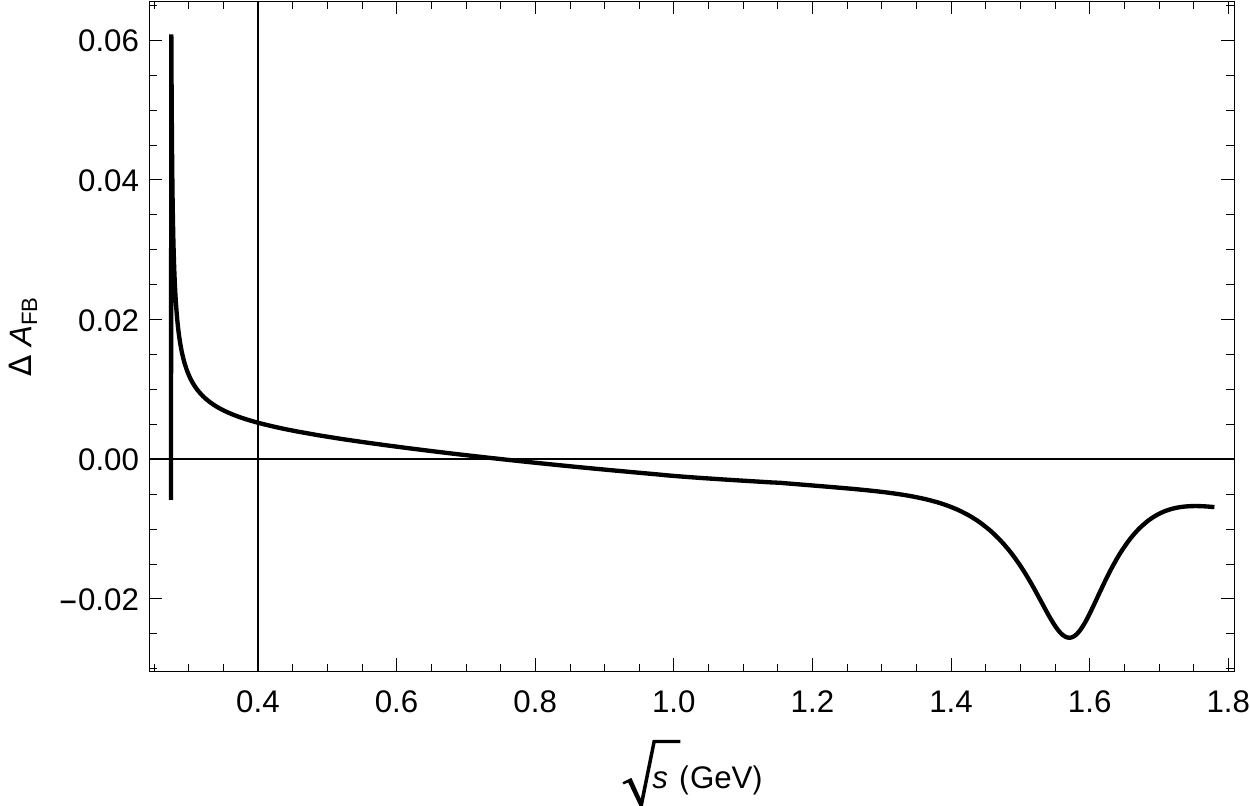}	
	\includegraphics[width=7cm]{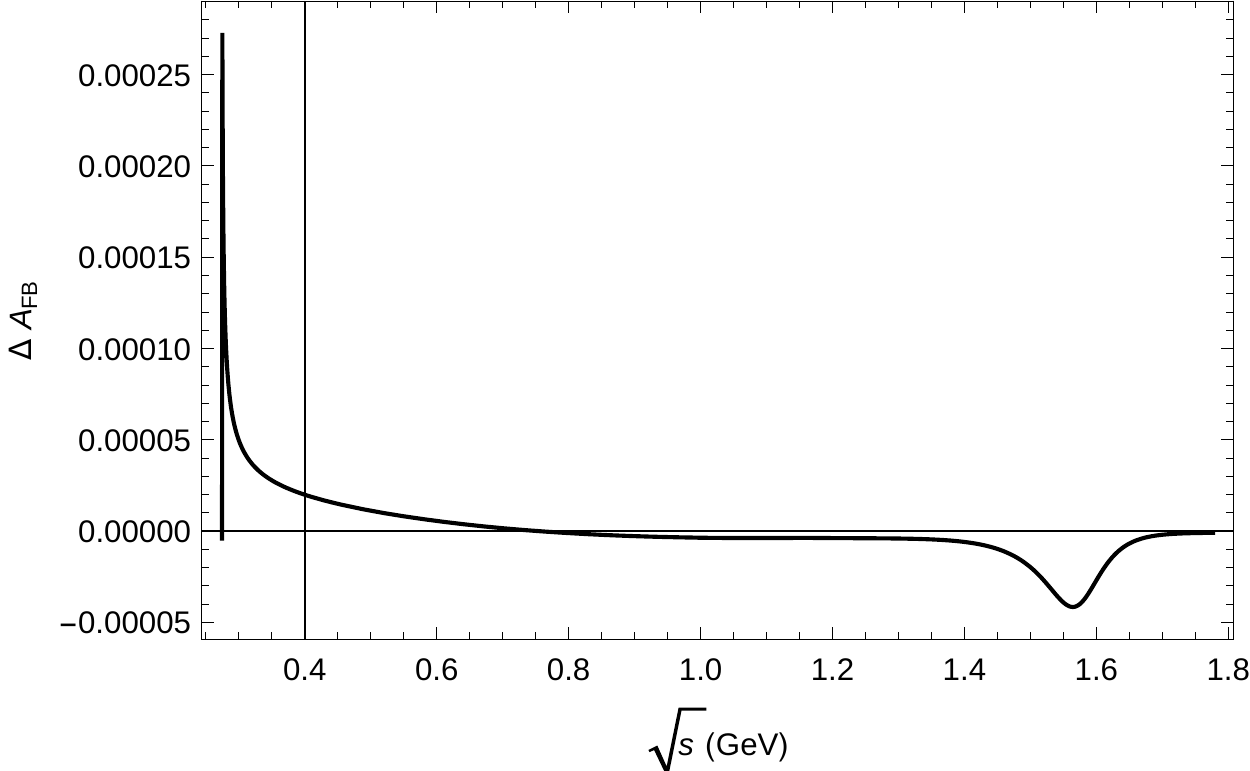}	
	\centering			
	\caption{Normalized difference with respect to the SM for the forward-backward asymmetry ($\Delta A_{FB}$) in the case of scalar interactions (left plot, with $\hat{\epsilon}_S=0.008,\hat{\epsilon}_T=0$) and tensor interactions (right plot, $\hat{\epsilon}_T=-0.001,\hat{\epsilon}_S=0$).}\label{Obs:fig8} 
\end{figure}

The forward-backward asymmetry is defined \cite{Descotes} by
\begin{equation}\label{FBA:eq1}
\mathcal{A}_{\pi\pi}(s)=\frac{\int_0^1 d\cos\theta \frac{d^2\Gamma}{dsd\cos\theta}-\int_{-1}^0 d\cos\theta \frac{d^2\Gamma}{dsd\cos\theta}}{\int_0^1 d\cos\theta \frac{d^2\Gamma}{dsd\cos\theta}+\int_{-1}^0 d\cos\theta \frac{d^2\Gamma}{dsd\cos\theta}}.
\end{equation}
We can obtain it for $\tau^-\to\pi^-\pi^0\nu_\tau$ decays plugging in eq. (\ref{a2:eq4}) into eq. (\ref{FBA:eq1}) and integrating upon the $\cos \theta$ variable,
\begin{equation}\begin{split}\label{AFB}
\mathcal{A}_{\pi\pi}(s)&=\frac{-3C_S\sqrt{\lambda\left(s,m_{\pi^-}^2,m_{\pi^0}^2\right)}}{2s^2\left[X_{VA}+\hat{\epsilon}_SX_S+\hat{\epsilon}_TX_T+\hat{\epsilon}_S^2X_{S^2}+\hat{\epsilon}_T^2X_{T^2}\right]}\left(1+\frac{s\hat{\epsilon}_S}{m_\tau (m_d-m_u)}\right)\Delta_{\pi^-\pi^0}\\
&\qquad\qquad\times \left\lbrace C_V \mathrm{Re}[F_0(s)F_+^*(s)]+\frac{2s\,\hat{\epsilon}_T}{m_\tau}\mathrm{Re}[F_T(s)F_0^*(s)]\right\rbrace,
\end{split}\end{equation}
where, again, the SM forward-backward asymmetry is recovered for $\epsilon_R=\epsilon_L=\hat{\epsilon}_S=\hat{\epsilon}_T=0$. This reference case is plotted in figure \ref{As:fig},  which agrees with the prediction in ref. \cite{Descotes} (this asymmetry was first studied in ref. \cite{Gao:2004gp}). This observable is plotted in fig. \ref{Obs:fig7} for an unrealistically large value of $\hat{\epsilon}_S$, for which there is a large deviation with respect to the SM case. Since such large departures disappear for reasonable values of $\hat{\epsilon}_{S,T}$, 
in order to enhance the sensitivity to new physics effects, we define the observable (odd under $\hat{\epsilon}_S\leftrightarrow -\hat{\epsilon}_S$)
\begin{equation}
 \Delta A_{FB}=A_{FB}(s,\hat{\epsilon}_S,\hat{\epsilon}_T)-A_{FB}(s,0,0),
\end{equation}
which is plotted in figs. \ref{Obs:fig8}. Even by using this observable it does not seem possible to evidence non-vanishing $\hat{\epsilon}_{S,T}$ using the forward-backward asymmetry.

As advanced before, $\mathcal{A}_{\pi\pi}(s)$ in eq. (\ref{AFB}) is a good observable for finding non-standard scalar interactions: despite its numerator is suppressed by the small value of $\Delta_{\pi^-\pi^0}$, its denominator is further suppressed by the dependence of $X_{S^2}$ on $\Delta_{\pi^-\pi^0}^2$, which enhances the sensitivity of this forward-backward asymmetry to scalar contributions. However, as just observed, if the strict limits on $|\hat{\epsilon}_S|$ obtained in other low-energy processes are applied, even $\mathcal{A}_{\pi\pi}(s)$ happens to be unable of evidencing this kind of NP contributions.

\subsection{Limits on $\hat{\epsilon}_S$ and $\hat{\epsilon}_T$}\label{limits}

\begin{figure}[h!]
\includegraphics[width=7cm]{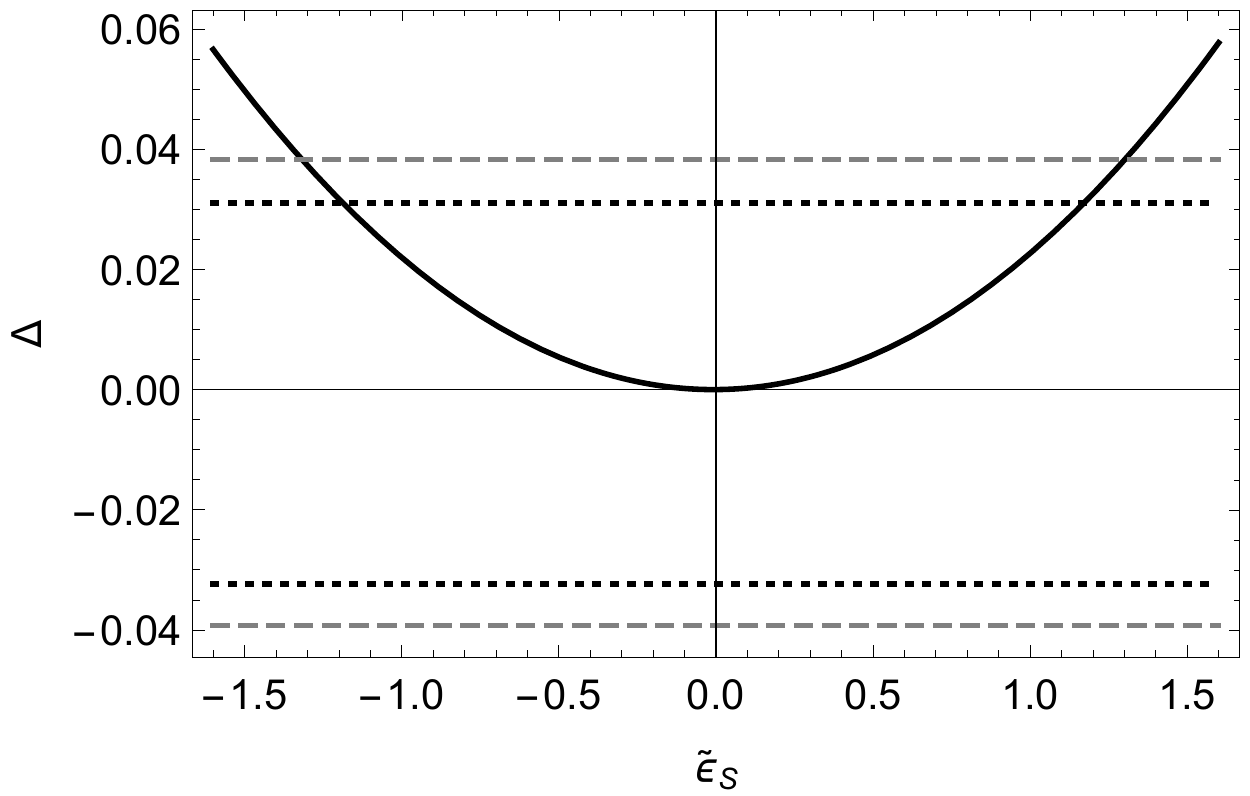}
\includegraphics[width=7cm]{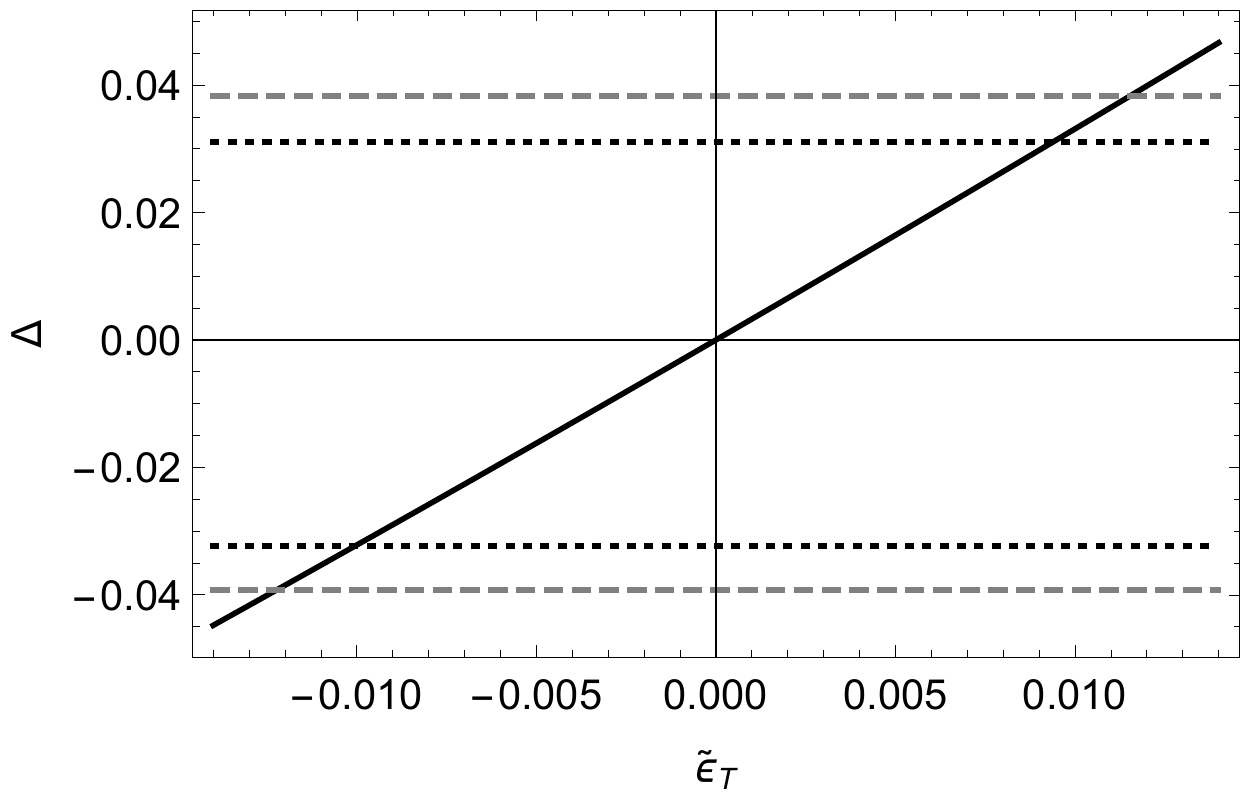}		
\centering			
\caption{$\Delta$ as a function of $\hat{\epsilon}_S$ for $\hat{\epsilon}_T=0$ (left-hand) and $\hat{\epsilon}_T$ for $\hat{\epsilon}_S=0$ (right-hand) for $\tau^-\to\pi^-\pi^0\nu_\tau$ decays. Horizontal lines represent the values of $\Delta$ according to the current measurement and theory error (at three standard deviations) of the branching ratio (dashed line) and the hypothetical case of this value being measured by Belle-II with three times reduced error (dotted line).}\label{Obs:fig9} 
\end{figure}

The $\tau^-\to\pi^-\pi^0\nu_\tau$ decay width can be obtained integrating the invariant mass distribution, using the expressions for the form factors \cite{Descotes, Dumm}. Since the total decay width depends on the effective couplings, this process branching ratio sets bounds on $\hat{\epsilon}_S$ and $\hat{\epsilon}_T$. For that, we compare the decay rate ($\Gamma$) for $\tau^-\to\pi^-\pi^0\nu_\tau$ in the presence of non-vanishing NP effective couplings with respect to the one ($\Gamma^0$) obtained by neglecting them (SM case). Using the best fit results of case III in ref.~\cite{Dumm}, we obtain a value of $\Gamma_0$ which corresponds to the branching ratio $(25.53\pm0.24)\%$, in excellent agreement with the PDG value of $(25.49\pm0.09)\%$. 
Integrating eq. (\ref{Dr:eq1}) we get the relative shift produced by NP contributions as follows
\begin{equation}\label{RD:eq1}
\Delta\equiv \frac{\Gamma-\Gamma^0}{\Gamma^0}=\alpha \hat{\epsilon}_S+\beta \hat{\epsilon}_T+\gamma \hat{\epsilon}_S^2+\delta \hat{\epsilon}_T^2\,,
\end{equation}
for whose coefficients we get: $\alpha=3.5\times10^{-4}$, $\beta=3.3^{+0.6}_{-0.4}$, $\gamma=2.2\times10^{-2}$, $\delta=4.7^{+2.0}_{-1.0}$. The relative error of the coefficients $\alpha$ and $\gamma$ due to our uncertainties is $\leq2\%$. Eq. (\ref{RD:eq1}) is a quadratic function of the effective scalar and tensor couplings, which can be used to explore the sensitivity of $\tau^-\to\pi^-\pi^0\nu_\tau$ decays to non-standard scalar and tensor interactions. We will do this in two steps. Firstly, we can make the analysis for one vanishing and one non-vanishing coupling. This is shown in figure \ref{Obs:fig9} where we represent with horizontal lines the current experimental limits on $\Delta$ (at three standard deviations) and use eq. (\ref{RD:eq1}) to translate this information into bounds for $\hat{\epsilon}_S$ and $\hat{\epsilon}_T$. According to this procedure, we get the following constraint $-1.33\leq \hat{\epsilon}_S\leq1.31$ with $\hat{\epsilon}_T=0$ and $[-0.79,-0.57]\cup[-1.4,1.3]\cdot10^{-2}$ as the allowed region for $\hat{\epsilon}_T$ with $\hat{\epsilon}_S=0$ (at three standard deviations). The previous results were used to estimate the values of $\hat{\epsilon}_S$ and $\hat{\epsilon}_T$ which were employed in the preceding subsections: $\hat{\epsilon}_S\sim1.31$ and $\hat{\epsilon}_T\sim-0.014$~\footnote{The value $\hat{\epsilon}_T\sim-0.001$ could seem a bit too small, compared to the intervals just given. However, we will see later in this section that the fits to the di-pion mass spectrum justify such an estimate.}. The dotted lines illustrate how the limits would evolve for an error reduced by a factor three, which could be achieved at Belle-II (the theory error is not assumed to decrease in this exercise).

\begin{figure}[t]
\includegraphics[width=7cm]{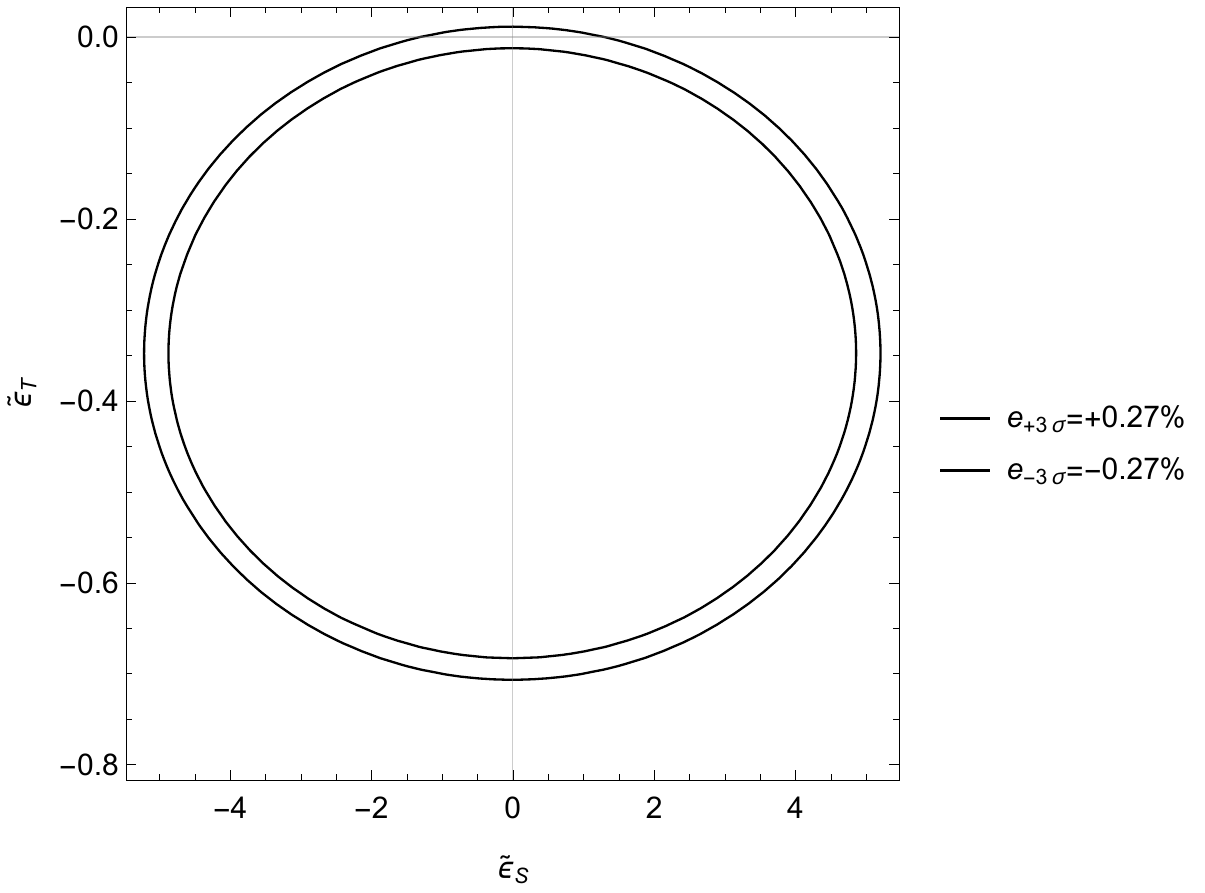}
\includegraphics[width=7cm]{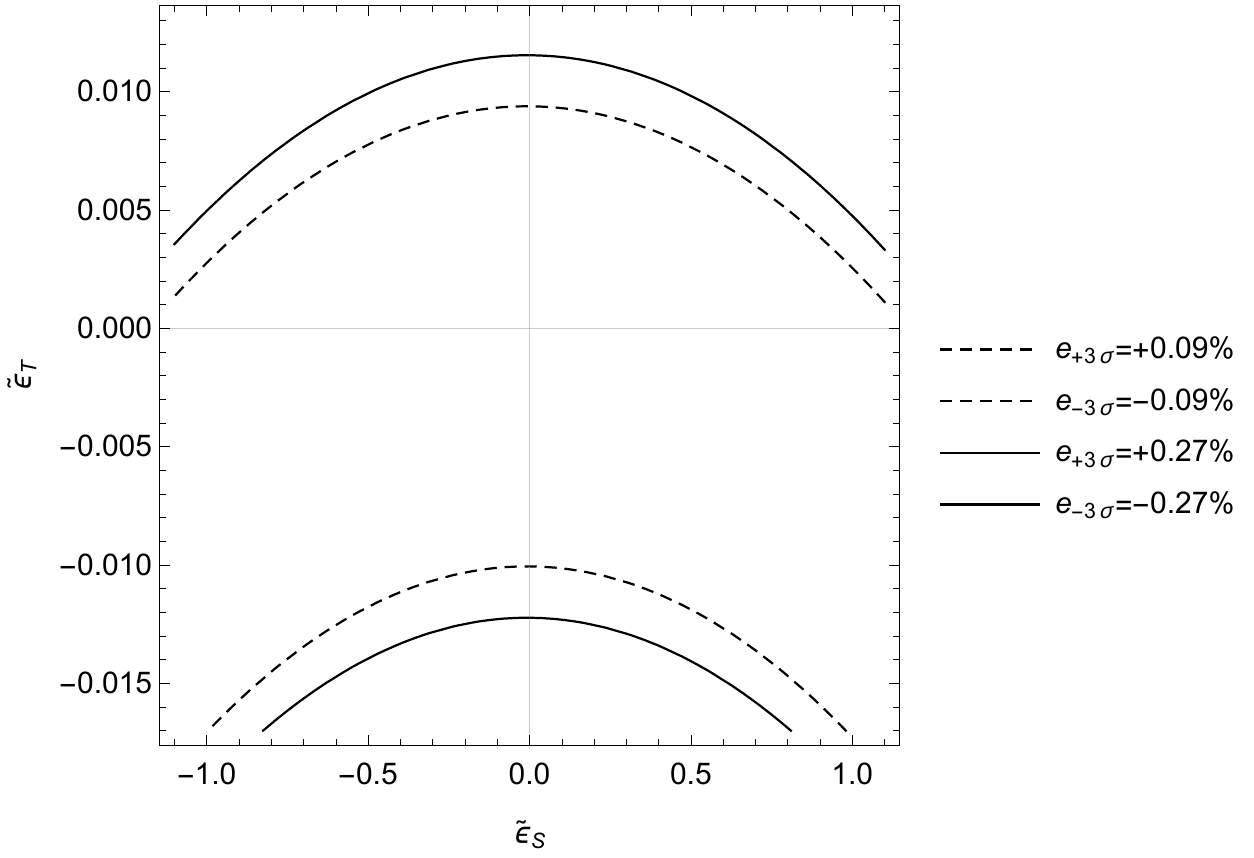}		
\centering			
\caption{Constraints on the scalar and tensor couplings obtained from $\Delta (\tau^-\to\pi^-\pi^0\nu_\tau)$ using the Belle measurement and the theory uncertainty (at three standard deviations) of the branching ratio. The left-hand plot shows the constraints obtained from current data. On the right-hand plot we show a magnification of the top part of this ellipse, where the solid line represents the upper limit on $\hat{\epsilon}_S$ and $\hat{\epsilon}_T$, while the dashed lines intend to illustrate the effect of a possible threefold improvement in the measurement at the Belle-II experiment.}\label{Obs:fig10}
\end{figure}

Then, we can also fix joint constraints on the scalar and tensor effective interactions assuming both $\hat{\epsilon}_S$ and $\hat{\epsilon}_T$ non-vanishing and using again eq. (\ref{RD:eq1}) as before. This result is shown in figure \ref{Obs:fig10}, where the limits on the scalar and tensor couplings are contained inside an ellipse in the $\hat{\epsilon}_S-\hat{\epsilon}_T$ plane. As a rough estimate of the possible impact of Belle-II data we repeat the exercise of assuming a threefold error improvement with respect to Belle-I. The dashed lines of the figure \ref{Obs:fig10} (right panel) are illustrative of this effect.

Table \ref{Obs:tab1} summarizes the constraints on the scalar and tensor effective couplings that can be obtained (at three standard deviations) from the Belle measurement of the branching ratio for $\tau^-\to\pi^-\pi^0\nu_\tau$ decays (including theory errors). The bottom part of table \ref{Obs:tab1} illustrates the bounds that could be achieved with a threefold reduction of the uncertainty at Belle-II.

\begin{table}[htbp]
\begin{center}
\begin{tabular}{|c|c|c|c|c|}
\hline
$\Delta$ limits & $\hat{\epsilon}_S\,(\hat{\epsilon}_T=0)$ & $\hat{\epsilon}_T\,(\hat{\epsilon}_S=0)$ & $\hat{\epsilon}_S$ & $\hat{\epsilon}_T$ \\
\hline
\hline
%$\pi^-\pi^0$ & & & & \\
Belle & $\left[-1.33,1.31\right]$ & \multicolumn{1}{p{2.8cm}|}{\centering $\left[-0.79,-0.57\right]\cup$\\ $\left[-1.4,1.3\right]\cdot 10^{-2}$} & $\left[-5.2,5.2\right]$ & $\left[-0.79,0.013\right]$\\ 
\multicolumn{1}{|p{2.8cm}|}{\centering3-fold improved\\ measurement} & $\left[-1.20,1.18\right]$ & \multicolumn{1}{p{3.3cm}|}{\centering $\left[-0.79,-0.57\right]\cup$\\ $\left[-1.1,1.1\right]\cdot 10^{-2}$} & $\left[-5.1,5.1\right]$ & $\left[-0.78,0.011\right]$\\
\hline
\end{tabular}
\caption{Constraints on the scalar and tensor couplings obtained (at three standard deviations) through the limits on the current branching ratio measurements and the hypothetical case where this value be measured by Belle II with a three times smaller error. Theory errors are included.}
\label{Obs:tab1}
\end{center}
\end{table}

Next we consider fits to the data reported by Belle \cite{Belle} for the normalized spectrum $(1/N_{\pi\pi})(dN_{\pi\pi}/ds)$ and integrated branching ratio using the function \footnote{All discussed uncertainties are considered in our fits.}
\begin{equation}
\frac{1}{\Gamma(\hat{\epsilon}_S,\hat{\epsilon}_T)}\frac{d\Gamma(s,\hat{\epsilon}_S,\hat{\epsilon}_T)}{ds}\,.
\end{equation}
When fitting $\hat{\epsilon}_S$ and $\hat{\epsilon}_T$ to Belle data in order to search for non-standard interactions, we are assuming that our description of $\Gamma_0$ (based on ref.~\cite{Dumm}) is a reliable estimate of the corresponding SM prediction (including theoretical uncertainties). Thus, we examine whether it is possible or not to improve the agreement of the SM prediction with data by means of non-vanishing new physics scalar or tensor interactions.

% In this way, we found the following constraints at one-sigma: $0.25\leq\vert \hat{\epsilon}_S\vert\leq0.52$ for the scalar coupling, and $0\leq|\hat{\epsilon}_T|\leq0.31\times 10^{-2}$ for the tensor coupling (the uncertainty on the sign and magnitude of $\Lambda_2$ was taken into account). %, which shows a moderate $1.5$ sigma discrepancy with respect to the SM, due to charged current tensor interactions that should be checked with more precise measurements of these decays and scrutinizing $F_+(s)$, hopefully with improved knowledge on $\Lambda_2$.
% As expected, the constraints on the scalar coupling are not competitive with refs. \cite{a24, a25, E}, because of the suppression of the scalar form factor by $\Delta_{\pi^-\pi^0}$. Considering this last feature,
If both $\hat{\epsilon}_S$ and $\hat{\epsilon}_T$ are fitted, bounds of order one on $\hat{\epsilon}_S$ and of order $0.1$ on $\hat{\epsilon}_T$ are obtained. Because of this unrealistic bounds for $\hat{\epsilon}_S$, which hinder the extraction of $\hat{\epsilon}_T$, in our reference fits we restrict $\vert \hat{\epsilon}_S\vert<0.8\times 10^{-2}$ \cite{a24, a25} and fit only $\hat{\epsilon}_T$. In this case we find $\hat{\epsilon}_T=\left(-1.3^{+1.5}_{-2.2}\right)\cdot10^{-3}$, which shows a small preference ($0.9$ sigma) for charged current tensor interactions. We believe, however, that it is interesting to check this conclusion with more precise measurements of these decays and scrutinizing $F_+(s)$, hopefully with improved knowledge on the inelastic effects on $F_T(s)$.

A caveat is, of course, in order: although chiral symmetry (at low energies) and the use of dispersion relations together with precise measurements (especially useful outside the $\chi PT$ regime of applicability) makes us confident on our knowledge of the vector two-pion form factor, $F_+(s)$, one should be very cautious before claiming evidence for NP from this type of analysis \footnote{In the case of $\tau^-\to\pi^- (\eta/\eta') \nu_\tau$ decays \cite{E} this would be noticeably more difficult: although the hadronization of the vector current is given again in terms of the precisely-known two-pion vector form factor, the dominant scalar contribution is subject to large uncertainties still \cite{Escribano:2016ntp}.}. Provided a hint for an anomaly appears, different investigations should be performed to test it: it may be worth considering a dispersive coupled-channel analysis of the two-pion and two-kaon vector form factors \cite{Guerrero:1998ei, Oller:2000ug, GomezNicola:2001as, Wilson:2015dqa}, one should analyze along these lines the compatibility between the $F_+(s)$ form factor measured by Belle and the $L=1=I$ $\pi\pi$ scattering amplitude...

We can finally compare the constraints in tables \ref{Obs:tab1} and the best fit results to the di-pion spectrum with those obtained in ref. \cite{a24}. For this, we need to assume lepton universality because our decays involve the tau lepton, while their analysis involves electron and muon flavors. According to refs. \cite{a24, a25, E}, it is clear $\tau^-\to\pi^-\pi^0\nu_\tau$ decays cannot be competitive setting constraints on the non-standard scalar interactions. Our three sigma upper limit (using current data) is $\vert \hat{\epsilon}_S\vert<1.3$ while the limit from the radiative pion decays $\pi\to e\nu\gamma$ is $\vert \hat{\epsilon}_S\vert<0.8\times 10^{-2}$ (at $90\%$ C.L.). Conversely, our %limit
best fit result, $\hat{\epsilon}_T=(-1.3^{+1.5}_{-2.2})\cdot10^{-3}$, is competitive in the case of tensor interactions %. T
since the limit reported in \cite{a24, a25} is $\vert \hat{\epsilon}_T\vert<0.1\times 10^{-2}$ (at $90\%$ C.L.). Notwithstanding, we find that the measured branching ratio only limits 
%and we get
$\hat{\epsilon}_T\in\left[-0.79,-0.57\right]\cup$ $\left[-1.4,1.3\right]\cdot 10^{-2}$ (at three sigma), which is not competitive with the previous value. Our results in this work and in ref.~\cite{E} are compatible with those in ref.~\cite{Cirigliano:2018dyk} (which also analyze semileptonic tau decays in this context): $\hat{\epsilon}_S=(-0.6\pm1.5)\cdot 10^{-2}$, $\hat{\epsilon}_T=(-0.04\pm0.46)\cdot 10^{-2}$. It must be noted that the analysis in ref.~\cite{E} does not include theory errors, which explains the smaller uncertainties quoted therein for $\hat{\epsilon}_S$. In this work, our bounds using only the measured branching ratio are less restrictive than those in ref.~\cite{Cirigliano:2018dyk}, and we can only achieve stronger limits with our fit to both the branching ratio and spectrum (using the error band for $\Gamma_0$ obtained in Ref. \cite{Dumm} and restricting $\vert \hat{\epsilon}_S\vert\lesssim1\times 10^{-2}$). In the light of more precise and diverse measurements of the $\tau^-\to\pi^-\pi^0\nu_\tau$ decays, improved theory analysis shall be needed to pursue cornering new physics with these decays.
 
%Finally, we collect in tables \ref{Obs:tab3} and \ref{Obs:tab4} the same results as in tables \ref{Obs:tab1} and \ref{Obs:tab2} but now using $\Lambda_2=\Lambda_2^{min}$. In this case, the hint for tensor charged current interactions is reduced from $\sim2.8$ to $\sim1.7$ sigma. Unfortunately, data cannot help us fix $\Lambda_2$. Apparently, only a lattice evaluation of the relevant matrix element -eq. (\ref{ad:3})- at the origin, could shed light on this issue, of utmost importance for restricting charged-current tensor interactions using  $\tau^-\to\pi^-\pi^0\nu_\tau$ decays data.

%\begin{table}[htbp]
%	\begin{center}
%		\begin{tabular}{|c|c|c|c|}
%			\hline
%			Fit & $\chi^2/d.o.f$ & $\hat{\epsilon}_S$ & $\hat{\epsilon}_T$ \\
%			\hline
%			\hline
%			& & & \\
%			Belle & $1.3$ & $\vert\hat{\epsilon}_S\vert < 0.28$ & $\left(1.0^{+0.2}_{-0.6}\right)\times 10^{-2}$ \\
%			\hline
%		\end{tabular}
%		\caption{Best fit results for $\hat{\epsilon}_S$ and $\hat{\epsilon}_T$ using $\Lambda_2=\Lambda_2^{min}$.}
%		\label{Obs:tab4}
%	\end{center}
%\end{table}

\section{SUMMARY AND CONCLUSIONS}\label{sec6}
We have considered the $\tau^-\to\pi^-\pi^0\nu_\tau$ decays in the presence of generic New Physics effective interactions up to dimension-six operators, assuming left-handed neutrinos and that the new dynamics scale is in the multi-TeV range. Within this setting, we have paid particular attention to the hadron matrix elements, which are needed SM inputs in order to set bounds on the non-standard scalar and tensor couplings, $\hat{\epsilon}_S$ and $\hat{\epsilon}_T$, respectively (we recall that it is not possible to restrict spin-one non-standard interactions in the considered processes). For this, we have employed previous results using dispersion relations for the scalar \cite{Descotes}, vector \cite{Dumm} and tensor \cite{Cirigliano:2017tqn} form factors implementing the known chiral constraints at low energies and QCD asymptotics at short distances, according to data. For the tensor form factor, since no experimental information is available, we have pursued a purely theoretical determination of its leading chiral behaviour using Chiral Perturbation Theory. In this work, we improved over our previous treatment of the tensor form factor where only leading-order chiral predictions were considered and unitarity constraints were ignored \cite{E}, motivated here by the fact that di-pion tau decays constitute an excellent arena to set competitive limits on $\hat{\epsilon}_T$. Lattice QCD results \cite{Baum:2011rm} allowed determining the only leading low-energy constant of the tensor form factor, permitting a direct access to $\hat{\epsilon}_T$.

 Within this framework, we have set bounds on $\hat{\epsilon}_S$ and $\hat{\epsilon}_T$ using the measured Belle branching ratio, through our observable $\Delta$. This procedure yields quite competitive limits with the world-best bounds for the tensor case (that we have thus used in the remaining analysis), but quite poor (unrealistic assuming some reasonable approximate lepton universality holds for them) in the scalar case, which is a consequence of its suppression in all considered observables (but the forward-backward asymmetry) by the tiny difference between charged and neutral pion masses squared. Because of this feature, we have assumed $\hat{\epsilon}_S$ limits similar to those obtained in light quark beta and $\tau^- \to \pi^- (\eta/\eta') \nu_\tau$ decays in the remaining analysis.
 
  As a result of our study, it turns out that Dalitz plot distributions (both in the Mandelstam variables $s$ and $t$ and also replacing $t$ by the angle between the two charged particles) are not very sensitive to non-zero realistic values of $\hat{\epsilon}_S$ and $\hat{\epsilon}_T$, as it also happens with the forward-backward asymmetry. Apparently, the hadronic invariant mass distribution is not sensitive either to charged-current tensor interactions. However, a fit to Belle data on this observable (limiting $\vert \hat{\epsilon}_S\vert\lesssim1\times 10^{-2}$ and with $\Gamma_0$ fixed -within errors- previously) hints for a slight preference for non-zero $\hat{\epsilon}_T$. Therefore, it is very worth measuring with extreme precision the di-pion invariant mass distribution in $\tau^-\to\pi^-\pi^0\nu_\tau$ decays at Belle-II, as it will serve to further restrict $\hat{\epsilon}_T$ and this way offer complementary information to other low-energy processes in the searches for non-standard charged current interactions. This effort would need to come together with both a tight scrutiny of the dominant vector form factor SM prediction and measurements of Dalitz distributions and forward-backward asymmetry.

\section*{ACKNOWLEDGEMENTS}
This work has been partially funded by Conacyt: the support of project 250628, as well as the scholarship during J. A. M. Ms. Sc. are acknowledged. We are indebted to Bastian Kubis for his very helpful suggestions to improve the hadronization of the tensor current with respect to the first draft of the paper.

\section*{APPENDIX: $F_T(s)$ including resonances as explicit degrees of freedom}
We show in this appendix that it is not convenient to build $F_T(s)/F_T(0)$ including resonances as explicit degrees of freedom.

As we will see, the tensor current couples to the $J^{PC}=1^{--}$ and $J^{PC}=1^{+-}$ resonances, but the contribution of the second tower of resonances is suppressed in the processes under consideration. This can be seen phenomenologically, since the $b_1(1235)$ resonance (which shares all quantum numbers with the $\rho(770)$ meson but has opposed parity) %decays only into four pions and $\bar{K}K\pi$, but 
is not known to couple to the two-pion system (precisely because of parity $b_1$ cannot decay into two pseudoscalars, though it could be exchanged in meson-meson scattering, but  $\pi\pi$ scattering data do not show any hint for exchange of the $b_1$ meson)%. This can indeed be understood from the fact that $b_1\to\pi\pi$ decays should be zero in the very approximate isospin symmetry limit \footnote{We thank Bastian Kubis for pointing this to us.}
. Therefore, the $\rho(770)$ is the lightest resonance whose exchange provides an energy-dependence to $F_T$, increasing its effect and allowing us to set more restrictive bounds on $\hat{\epsilon}_T$ (we neglect the contributions from $\rho$ excitations in this study). 

We shall now discuss the chiral couplings of meson resonances to the pseudoscalar Goldstone fields in the presence of tensor currents. We use the antisymmetric tensor representation \cite{Ecker, Ecker:1989yg} in order to describe the relevant spin-one degrees of freedom. To determine the resonance exchange contributions to the 
$\tau^-\to\pi^-\pi^0\nu_\tau$ decays (or to the effective chiral Lagrangian) we need the lowest order operators in the chiral expansion which are linear in the resonance fields. 
Using the $P$ and $C$ transformation properties of given $J^{PC}$ resonance fields: $V(1^{--})$, $A(1^{++})$, $S(0^{++})$, $P(0^{-+})$ (see Table 2 in ref. \cite{Ecker}), and $H(1^{+-})$ and $T(2^{++})$ (see ref. \cite{Ecker2}), we can, for the first time, construct the $R\chi T$ Lagrangian linear in resonance fields and coupled to the tensor source of lowest chiral 
order, which has the following two pieces:

\begin{subequations}
	\begin{align}
	\mathcal{L}[V(1^{--})]&=F_V^T M_V\left\langle V_{\mu\nu}t_+^{\mu\nu}\right\rangle,\\
	\mathcal{L}[H(1^{+-})]&=iF^T_H M_H \left\langle H_{\mu\nu}t_-^{\mu\nu}\right\rangle.
	\end{align}
\end{subequations}

In the following, we neglect the effect of the latter operator (assuming $F^T_H$ negligible) because of the seemingly small $b_1\pi\pi$ coupling commented above. A straightforward computation of the contribution of the former operator to the relevant hadronic matrix element yields

\begin{equation}
\langle \pi^0\pi^- \vert \bar{d}\sigma^{\mu\nu}u \vert 0\rangle=iF_T(s) \left(p^\mu_{\pi^0}p^\nu_{\pi^-}-p^\mu_{\pi^-}p^\nu_{\pi^0}\right)\,,\label{FTFF}
\end{equation}
where
\begin{equation}
F_T(s)=\frac{\sqrt{2}\Lambda_2}{F^2}\left[1+\frac{G_VF_V^T}{\Lambda_2}\frac{M_\rho}{M_\rho^2-s}\right]\,,\label{FTFFs}
\end{equation}
in which the operator $\frac{iG_V}{\sqrt{2}}\left\langle V_{\mu\nu} u^\mu u^\nu\right\rangle$ \cite{Ecker} was used in order to obtain the $\rho\pi\pi$ coupling.

Eq.~(\ref{FTFFs}) depends on three \textit{a priori} unknown couplings. Fortunately, short-distance QCD properties can shed light on their values, as we explain next. First, it is known from the analysis of two-point correlators within $R\chi T$ that $G_V=F/\sqrt{2}$ \cite{Ecker} (also $F_V=\sqrt{2}F$, which is used next). The large-$N_C$ 
asymptotic analysis of $\langle VV \rangle$, $\langle TT \rangle$ and $\langle VT \rangle$ correlators determines $F_V^T/F_V=1/\sqrt{2}$ \cite{Cata}, in such a way that only $\Lambda_2$ remains unrestricted and eq.~(\ref{FTFFs}) simplifies to 
\begin{equation}
F_T(s)=\frac{\sqrt{2}\Lambda_2}{F^2}+\frac{M_\rho}{M_\rho^2-s}\,.\label{FTFFssimp}
\end{equation}

The $\rho$ meson contribution shifts the value of $F_T(0)$ by $\sim 65\%$, which is unphysical.

As in the case of the vector form factor, the $\rho$-propagator in eq.~(\ref{FTFFs}) is modified by the inclusion of the width $\Gamma_\rho(s)$ (proportional to the imaginary part of the corresponding loop contributions) and also by shifting the pole mass value (according to the real part of the loop contribution), as required by 
analyticity. Specifically,
\begin{equation}\label{analytic propagator}
(M_\rho^2-x)^{-1}\to\left\lbrace M_\rho^2\left(1+\frac{x}{96\pi^2F^2}\mathrm{Re} \left[A_\pi(x)+\frac{A_K(x)}{2}\right]\right)-x-i M_\rho \Gamma_\rho(x)\right\rbrace^{-1}\,,
\end{equation}
with
\begin{equation}\begin{split}
\Gamma_\rho(x) & =  \frac{M_\rho x}{96\pi F^2}\left[ \theta(x-4m_\pi^2) \sigma_\pi^3(x)+\theta(x-4m_K^2) \frac{\sigma_K^3(x)}{2}\right]\nonumber\\
& =  -\frac{M_\rho x}{96\pi^2F^2}\mathrm{Im} \left[ A\left(\frac{m_\pi^2}{x},\frac{m_\pi^2}{M_\rho^2}\right)+\frac{1}{2} A\left(\frac{m_K^2}{x},\frac{m_K^2}{M_\rho^2}\right)\right]
\end{split}\end{equation}
and ($A_P(x)$ is short for 
$A\left(\frac{m_P^2}{x},\frac{m_P^2}{M_\rho^2}\right)$)
\begin{equation}
\mathrm{Re}  A_P(x)\,=\,Log\frac{m_P^2}{M_\rho^2}+8\frac{m_P^2}{x}-\frac{5}{3}+\sigma_P^3(x)Log\Bigg|\frac{\sigma_P(x)+1}{\sigma_P(x)-1}\Bigg|\,,
\end{equation}
being $\sigma_P(x)=\sqrt{1-\frac{4m_P^2}{x}}$.

\begin{figure}[t]
	\includegraphics[width=7cm]{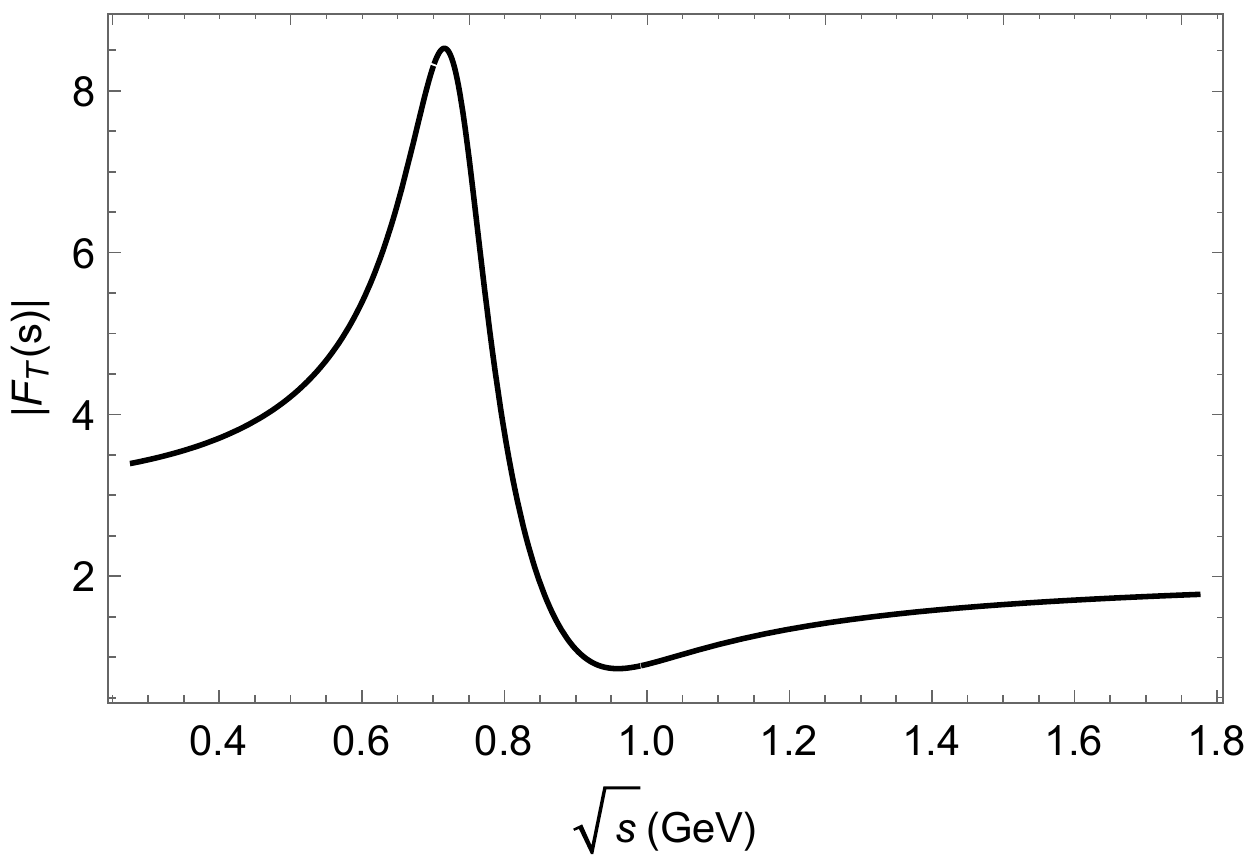}
	\includegraphics[width=7cm]{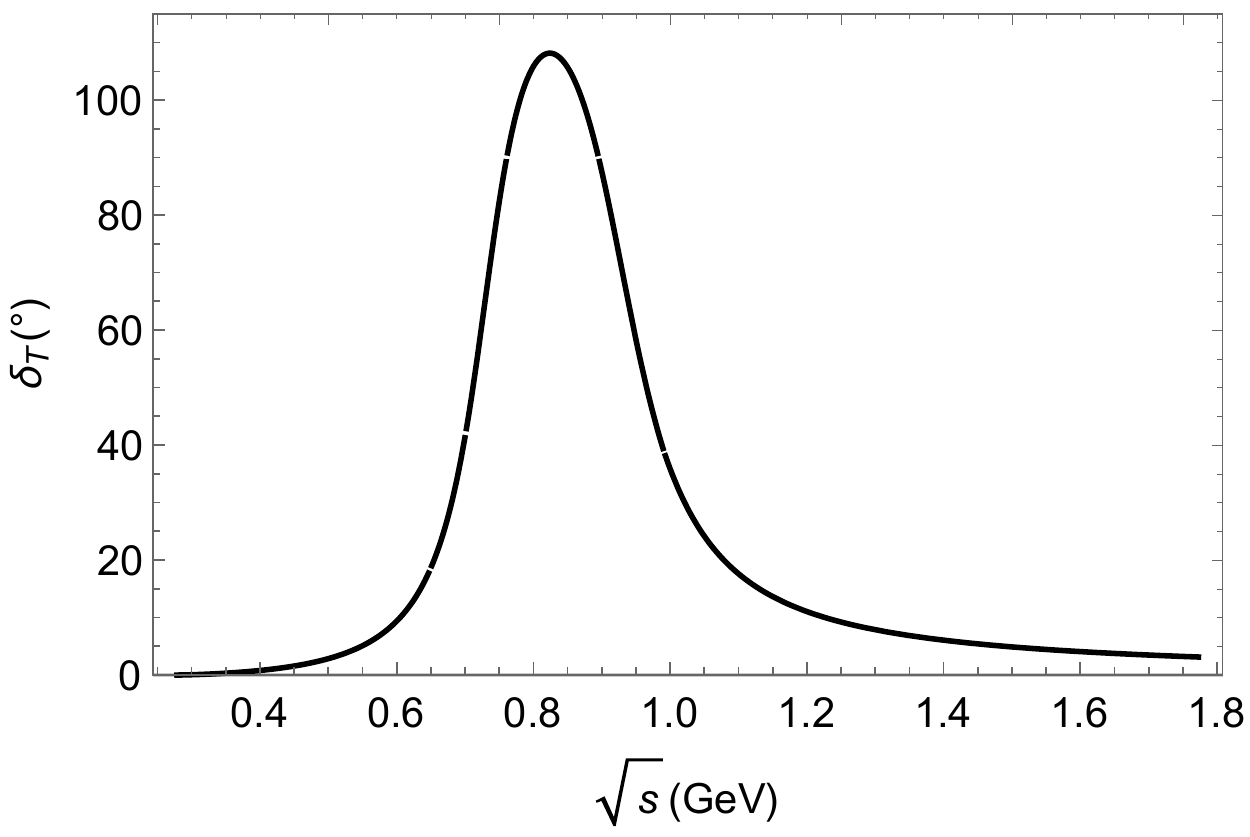}
	\centering
	\caption{Modulus and phase, $|F_T(s)|$ (left) and $\delta_T(s)$ (right), of the tensor form factor, $F_T(s)$, corresponding to the description explained in this appendix.}\label{PlotsFTs}
\end{figure}

The tensor form factor, $F_T(s)$, given by eq.~(\ref{FTFFssimp}), and using the substitution eq.~(\ref{analytic propagator}), is plotted in figure \ref{PlotsFTs} for $\Lambda_2=12$ MeV \cite{Baum:2011rm}. There, it is seen how the $\rho(770)$ meson contribution modifies the constant $\chi PT$ lowest-order result for $|F_T(s)|$. The form factor 
phase, $\delta_T(s)$, grows from zero to $\sim110^\circ$ for $0.85\leq\sqrt{s}\leq0.90$ GeV and decreases softly to zero for larger energies. Both $|F_T(s)|$ and $\delta_T(s)$ are influenced by the on-shell $\rho(770)$ meson width as expected, according to its value of $\sim145$ MeV.

At this point unitarity arguments may convince us that this description of $F_T(s)$ cannot be complete \footnote{We thank Bastian Kubis for pointing this to us.}. As explained in ref. \cite{Cirigliano:2017tqn}, the phase of $F_T(s)$ must coincide with the phase of $F_+(s)$ in the elastic region (in this paper this was shown for the tau decays into the $K\pi$ system, but it is completely analogous to the $\pi\pi$ one considered here). We briefly review the argument in what follows.

The unitarity relation for $F_+(s)$ can be written
\begin{equation}
\Im m F_+(s) = \sigma_\pi(s) F_+(s) (f_1^1(s))^* \theta(s-4m_\pi^2)\,,
\end{equation}
where $f_1^1(s)$ is the the corresponding partial wave in $\pi\pi$ scattering. The previous equation implies that, in the elastic region, $\delta_1^1(s)=\delta_+(s)$, which is again Watson's theorem. The crucial point is that an analogous unitarity relation holds for $F_T(s)$:
\begin{equation}
\Im m F_T(s) = \sigma_\pi(s) F_T(s) (f_1^1(s))^* \theta(s-4m_\pi^2)\,,\label{phaserelation}
\end{equation}
from which one can immediately derive that, in the elastic region, $\delta_T(s)=\delta_+(s)$, a feature that is not satisfied by our expression for $F_T(s)$ considered up to now (and it will not be satisfied for any value of $\Lambda_2$). This should not be understood as a failure of eq. ~(\ref{FTFFssimp}) (together with eq.~(\ref{analytic propagator})), but rather as a manifestation of its incompleteness. Indeed, the contributions from the next-to-leading order $\chi PT$ Lagrangian with tensor sources ($\mathcal{O}(p^6)$ in the chiral counting \cite{Mateu}) should provide with the needed energy-dependence to satisfy eq. (\ref{phaserelation}). However, since the number of such operators is 75 (plus 3 contact terms) even in the $SU(2)$ case \cite{Mateu}, we refrain from proceeding this way as any predictability would be lost.

\bibliographystyle{unsrt}
\bibliography{lib}
\addcontentsline{toc}{section}{References}

\end{document}